\documentclass[twocolumn]{aastex62}

\usepackage{amsmath}

\newcommand\target{MACS0416\_Y1}
\newcommand\oiii{[\textsc{O\,iii}]}
\newcommand\cii{[\textsc{C\,ii}]}
\newcommand\Loiii{L_\mathrm{[O\,III]}}
\newcommand\Lir{L_\mathrm{IR}}
\newcommand\Tdust{T_\mathrm{dust}}
\newcommand\Mdust{M_\mathrm{dust}}
\newcommand\Mstar{M_\mathrm{star}}

\newcommand\HST{\textit{HST}}
\newcommand\Hubble{\textit{Hubble}}
\newcommand\Spitzer{\textit{Spitzer}}
\newcommand\Herschel{\textit{Herschel}}
\newcommand\Gaia{\textit{Gaia}}

\received{2018 June 10}
\accepted{2019 January 31}

\shorttitle{\textsc{[O\,iii]} and Dust at $z = 8.312$}
\shortauthors{Tamura et al.}

\begin{document}

\title{Detection of the Far-infrared \textsc{[O\,iii]} and Dust Emission in a Galaxy at Redshift 8.312: \\Early Metal Enrichment in the Heart of the Reionization Era}

\correspondingauthor{Yoichi Tamura}
\email{ytamura@nagoya-u.jp}

\author[0000-0003-4807-8117]{Yoichi Tamura}
\affil{Division of Particle and Astrophysical Science, Graduate School of Science, Nagoya University, Nagoya 464-8602, Japan.}

\author{Ken Mawatari}
\affil{Department of Environmental Science and Technology, Faculty of Design Technology, Osaka Sangyo University, 3-1-1, Nakagaito, Daito, Osaka 574-8530, Japan.}
\affil{Institute for Cosmic Ray Research, The University of Tokyo, Kashiwa, Chiba 277-8582, Japan.}

\author[0000-0002-0898-4038]{Takuya Hashimoto}
\affil{Department of Environmental Science and Technology, Faculty of Design Technology, Osaka Sangyo University, 3-1-1, Nakagaito, Daito, Osaka 574-8530, Japan.}
\affil{National Astronomical Observatory of Japan, 2-21-1, Osawa, Mitaka, Tokyo 181-8588, Japan.}

\author[0000-0002-7779-8677]{Akio K.\ Inoue}
\affil{Department of Environmental Science and Technology, Faculty of Design Technology, Osaka Sangyo University, 3-1-1, Nakagaito, Daito, Osaka 574-8530, Japan.}

\author[0000-0003-1096-2636]{Erik Zackrisson}
\affil{Observational Astrophysics, Department of Physics and Astronomy, Uppsala University, Box 516, SE-751 20 Uppsala, Sweden}

\author[0000-0001-8415-7547]{Lise Christensen}
\affil{Dark Cosmology Centre, Niels Bohr Institute, University of Copenhagen, Juliane Maries Vej 30, DK-2100 Copenhagen, Denmark}

\author{Christian Binggeli}
\affil{Observational Astrophysics, Department of Physics and Astronomy, Uppsala University, Box 516, SE-751 20 Uppsala, Sweden}

\author{Yuichi Matsuda}
\affil{National Astronomical Observatory of Japan, 2-21-1, Osawa, Mitaka, Tokyo 181-8588, Japan.}
\affil{The Graduate University for Advanced Studies (SOKENDAI), 2-21-1, Osawa, Mitaka, Tokyo 181-8588, Japan.}

\author{Hiroshi Matsuo}
\affil{National Astronomical Observatory of Japan, 2-21-1, Osawa, Mitaka, Tokyo 181-8588, Japan.}
\affil{The Graduate University for Advanced Studies (SOKENDAI), 2-21-1, Osawa, Mitaka, Tokyo 181-8588, Japan.}

\author{Tsutomu T.\ Takeuchi}
\affil{Division of Particle and Astrophysical Science, Graduate School of Science, Nagoya University, Nagoya 464-8602, Japan.}

\author{Ryosuke S.\ Asano}
\affil{Division of Particle and Astrophysical Science, Graduate School of Science, Nagoya University, Nagoya 464-8602, Japan.}

\author{Kaho Sunaga}
\affil{Division of Particle and Astrophysical Science, Graduate School of Science, Nagoya University, Nagoya 464-8602, Japan.}

\author{Ikkoh Shimizu}
\affil{Department of Earth \& Space Science, Osaka University, 1-1, Machikaneyama, Toyonaka, Osaka 560-0043, Japan}

\author[0000-0003-0137-2490]{Takashi Okamoto}
\affil{Department of Cosmosciences, Graduates School of Science, Hokkaido University, N10W8, Kitaku, Sapporo 060-0810, Japan}

\author[0000-0001-7925-238X]{Naoki Yoshida}
\affil{Department of Physics, Graduate School of Science, The University of Tokyo, Tokyo 113-0033, Japan}
\affil{Kavli Institute for the Physics and Mathematics of the Universe (WPI), Todai Institutes for Advanced Study, The University of Tokyo, Kashiwa, Chiba 277-8583, Japan.}

\author[0000-0002-2419-3068]{Minju M.\ Lee}
\affil{Division of Particle and Astrophysical Science, Graduate School of Science, Nagoya University, Nagoya 464-8602, Japan.}
\affil{National Astronomical Observatory of Japan, 2-21-1, Osawa, Mitaka, Tokyo 181-8588, Japan.}

\author{Takatoshi Shibuya}
\affil{Department of Computer Science, Kitami Institute of Technology, 165 Koen-cho, Kitami, Hokkaido 090-8507, Japan}

\author[0000-0003-2247-3741]{Yoshiaki Taniguchi}
\affil{The Open University of Japan, 2-11 Wakaba, Mihama-ku, Chiba 261-8586, Japan}

\author[0000-0003-1937-0573]{Hideki Umehata}
\affil{The Open University of Japan, 2-11 Wakaba, Mihama-ku, Chiba 261-8586, Japan}
\affil{RIKEN Cluster for Pioneering Research, 2-1 Hirosawa, Wako-shi, Saitama 351-0198, Japan}
\affil{Institute of Astronomy, The University of Tokyo, 2-21-1, Osawa, Mitaka, Tokyo 181-0015, Japan}

\author[0000-0001-6469-8725]{Bunyo Hatsukade}
\affil{Institute of Astronomy, The University of Tokyo, 2-21-1, Osawa, Mitaka, Tokyo 181-0015, Japan}

\author[0000-0002-4052-2394]{Kotaro Kohno}
\affil{Institute of Astronomy, The University of Tokyo, 2-21-1, Osawa, Mitaka, Tokyo 181-0015, Japan}
\affil{Research Center for the Early Universe, Graduate School of Science, The University of Tokyo, Tokyo 113-0033, Japan}

\author[0000-0002-7675-5923]{Kazuaki Ota}
\affil{Kavli Institute for Cosmology, University of Cambridge, Madingley Road, Cambridge, CB3 0HA, UK}
\affil{Kyoto University Research Administration Office, Yoshida-Honmachi, Sakyo-ku, Kyoto 606-8501, Japan}




\begin{abstract}

We present the Atacama Large Millimeter/submillimeter Array (ALMA) detection of the \oiii{} 88~$\micron$ line and rest-frame 90~$\micron$ dust continuum emission in a $Y$-dropout Lyman break galaxy (LBG), \target{}, lying behind the Frontier Field cluster MACS J0416.1$-$2403.
This \oiii{} detection confirms the LBG with a spectroscopic redshift of $z = 8.3118 \pm 0.0003$, making this object one of the furthest galaxies ever identified spectroscopically.
The observed 850~$\micron$ flux density of $137 \pm 26$~$\mu$Jy corresponds to a de-lensed total infrared (IR) luminosity of $\Lir = (1.7 \pm 0.3) \times 10^{11}\,L_\sun$ if assuming a dust temperature of $T_\mathrm{dust} = 50$ K and an emissivity index of $\beta = 1.5$, yielding a large dust mass of $4 \times 10^6 M_\sun$.
The ultraviolet-to-far IR spectral energy distribution modeling where the \oiii{} emissivity model is incorporated suggests the presence of a
young ($\tau_\mathrm{age} \approx 4$~Myr), star-forming (SFR $\approx 60~M_{\sun}$~yr$^{-1}$), moderately metal-polluted ($Z \approx 0.2 Z_{\odot}$) stellar component with a mass of $M_\mathrm{star} = 3 \times 10^8 M_{\odot}$.
An analytic dust mass evolution model
with a single episode of star-formation does not reproduce the metallicity and dust mass in $\tau_\mathrm{age} \approx 4$~Myr, suggesting a pre-existing evolved stellar component with $M_\mathrm{star} \sim 3 \times 10^9~M_{\sun}$ and $\tau_\mathrm{age} \sim 0.3$~Gyr as the origin of the dust mass.

\end{abstract}




\section{Introduction} \label{sec:intro}

    How and when metal enrichment happened in the epoch of reionization (EoR) is one of the most fundamental questions in modern astronomy. Recent \textit{Planck} results suggest that the cosmic reionization occurred at an instantaneous reionization redshift of
    $z_\mathrm{re} = 7.68 \pm 0.79$ \citep{Planck18},
    and the latest \textit{Hubble Space Telescope} (\HST{}) surveys have revealed more than a hundred of candidate $z \gtrsim 8$ Lyman break galaxies \citep[LBGs, e.g.,][see \citealt{Stark16} for a review]{Bouwens15, Ishigaki18} out to $z = 11.1^{+0.08}_{-0.12}$ \citep{Oesch16}.
    Furthermore, based on the samples of $z \gtrsim 8$ LBGs, \citet{Oesch18} reported a strong evolution of the ultraviolet (UV) luminosity function by one order of magnitude from $z \sim 10$ to $\sim 8$, implying a rapid increase of the cosmic star-formation rate density by an order of magnitude within a very short time-scale ($\lesssim 200$~Myr).  It is likely that this steep evolution compared to lower-$z$ can be explained by the fast build-up of the dark matter halo mass function at $z > 8$ \citep{Oesch18}.

    However, it is still a challenge to characterize baryonic physics of the $z > 8$ galaxies.
    One of the major obstacles is that these LBGs are yet to be confirmed through spectroscopy; since the rest-frame UV continuum is typically not sufficiently bright for detection with current instruments, it is often assumed that the Lyman-$\alpha$ (Ly$\alpha$) emission line might be the best tool for spectroscopic confirmation.  A large amount of 8--10~m telescope time have been invested in Ly$\alpha$ searches for $z > 8$ candidates, but so far this resulted in only a few detections \citep[$z = 8.683$, 8.38, 9.11,][respectively]{Zitrin15, Laporte17, Hashimoto18}, likely indicating that the Ly$\alpha$ signal is substantially attenuated by the largely-neutral intergalactic medium at this epoch.

    Alternative UV indicators such as \textsc{C~iii]} $\lambda\lambda$1907,1909~\AA{} serve as a workhorse for redshift identification \citep[e.g.,][]{Zitrin15a, Stark15a}. These lines, in addition to the rest-frame optical oxygen and nitrogen lines, are also useful for characterizing the metal enrichment of interstellar medium (ISM) and stellar components of the galaxies \citep[e.g.,][]{Stark15a, Stark15b, Stark17, Mainali18}.
    These diagnostic lines will provide a unique insight into the physical properties of ionized gas in the $z > 8$ Universe when the \emph{James Webb Space Telescope} (\emph{JWST}) comes online, although their use is currently limited to bright galaxies in a certain redshift range, because most of the lines are intrinsically faint and/or are redshifted outside the wavelength range where the atmospheric transmittance is good for ground-based facilities.

    With this in mind, \citet{Inoue14} have investigated the potential use of redshifted nebular emission lines in the rest-frame far-infrared (FIR) in determining spectroscopic redshift of $z \sim 8$ galaxies.  The \oiii{} 88~$\micron$ line, which is often observed as the brightest FIR line in local \textsc{H~ii} regions \citep[e.g.,][]{Takami87, Kawada11}, can be used as an instantaneous tracer of massive star formation, since ionization of O$^{+} \rightarrow$ O$^{++}$ requires hard ($E >35.1$~eV) ionizing photons from hot, short-lived O-type stars.
    \citet{Inoue14} predicted the line fluxes from high-$z$ galaxies on the basis of a cosmological hydrodynamic simulation of galaxy formation \citep{Shimizu14} by incorporating an \oiii{} emission line model as a function of metallicity calibrated by \textit{ISO}, \textit{AKARI} and \Herschel{} observations of local galaxies \citep[][see also \citealt{Cormier15}]{Brauher08, Kawada11, Madden12, Madden13}.
    Since the metallicity of a typical galaxy with $H_{\rm 160} = 26$\,mag (AB) reaches $\sim 0.2$ $Z_{\odot}$ even at $z \gtrsim 8$, the \oiii{} line of such galaxies is as bright as 1--5~mJy, which is bright enough to be detected with existing submillimeter facilities, such as the Atacama Large Millimeter/submillimeter Array (ALMA).

    Indeed, it is becoming clear that galaxies at $z > 6$ are bright in \oiii{} 88~$\micron$ \citep{Inoue16, Carniani17, Laporte17, Marrone18, Hashimoto18, Hashimoto18b, Hashimoto18c, Walter18}.  The first detection in the EoR has been made for a $z = 7.212$ Ly$\alpha$ emitter, SXDF-NB1006-2 \citep{Inoue16}, in which only 2~hr integration of ALMA Band 8 was invested, implying the \oiii{} line as a promising tool to pin down the spectroscopic redshift even for $z > 8$ galaxies.
    More recently, two LBGs, A2744\_YD4 at $z = 8.38$ \citep{Laporte17} and MACS1149-JD1 at $z = 9.1096 \pm 0.0006$ \citep{Hashimoto18}, have been confirmed in \oiii{} at $4.0\sigma$ and $7.4\sigma$, respectively.  A2744\_YD4 was also detected in 850-$\micron$ continuum with $S_\mathrm{850\,\mu m} = 0.1$~mJy, suggesting the presence of a chemically-evolved ISM.  It should also be noted that MACS1149-JD1 was identified without any prior information of a redshift inferred from a spectral line, demonstrating the \oiii{} line as the redshift indicator complementing the role of the Ly$\alpha$ and other UV lines.

    In addition, the \oiii{} 88~$\micron$ flux places a unique constraint on the stellar and ISM properties, since the \oiii{} line is extinction-free and sensitive to the electron density, ionization parameter, and gas-phase oxygen abundance of the ionized media, which also depend on the global properties such as the star-formation rate (SFR) and stellar age.  \citet{Inoue16} have carried out comprehensive modeling of the UV-to-FIR spectral energy distribution (SED) of SXDF-NB1006-2, in which the \oiii{} flux and submillimeter continuum upper limits are taken into account. They found this to be a young ($< 30$~Myr) star-forming ($\approx 300~M_{\odot}$~yr$^{-1}$) galaxy with a somewhat high best-fitting metallicity of 0.05--1 $Z_{\odot}$.  In contrast, \citet{Hashimoto18} revealed that MACS1149-JD1 has a more evolved (290~Myr) stellar component with a metallicity of $0.2Z_{\sun}$, suggesting a formation redshift of $z_\mathrm{f} \approx 15$.
    The non-detection of dust continuum in both galaxies suggests low dust-to-metal mass ratios in their ISM \citep{Inoue16} compared to the Milky Way's value \citep[$\sim 0.5$, e.g.,][]{Inoue11b}, which could indicate that a substantial fraction of ISM metals is not stored in grains.

    In this paper, we report the detections of the \oiii{} 88~$\micron$ line and dust continuum in a modestly-magnified $Y$-dropout LBG, confirming the spectroscopic redshift to be $z = 8.3118 \pm 0.0003$, i.e., corresponding to an epoch when the age of the Universe was only 600 Myr.  This is one of the furthest galaxies ever identified spectroscopically by exploiting the brightness of the \oiii{} line.  This paper is organized as follows:
        \S~\ref{sec:target} explains how the target was selected.
        \S~\ref{sec:obs} describes the ALMA and VLT/X-shooter observations.
        \S~\ref{sec:result} demonstrates the detection of dust and \oiii{} emission in \target{}.
        In \S~\ref{sec:sed}, we perform an analysis of the SED to constrain the physical properties of \target{}.
        In \S~\ref{sec:discussion}, we discuss the model prediction of the dust mass, whereby we demonstrate that some parameter degeneracies obtained in the SED analysis can be resolved by incorporating a dust mass evolution model.
        Finally, our conclusions are presented in \S~\ref{sec:conclusion}.

    Throughout this paper, we adopt a concordance cosmology with $\Omega_\mathrm{m} = 0.3$, $\Omega_{\Lambda} = 0.7$ and $H_0 = 70$ km\,s$^{-1}$\,Mpc$^{-1}$.  An angular scale of $1''$ corresponds to the physical scale of 4.7~kpc at $z = 8.312$.  A redshift $z = 8.312$ corresponds to an age of the Universe of 0.60~Gyr.


    \begin{deluxetable*}{cccccc}[ht!]
    \tablecaption{The parameters of ALMA observations. \label{tab:log}}
    \tablecolumns{6}
    \tablewidth{0pt}
    \tablehead{
    \colhead{UT start time\tablenotemark{$\sharp$}} &
    \colhead{Baseline lengths} &
    \colhead{} &
    \colhead{Center frequency} & \colhead{Integration time} & \colhead{PWV} \\
    \colhead{(YYYY-MM-DD hh:mm:ss)} & \colhead{(m)} & \colhead{$N_\mathrm{ant}$\tablenotemark{$\dagger$}} &
    \colhead{(GHz)} & \colhead{(min)} & \colhead{(mm)}
    }
    \startdata
    2016-10-25 05:11:40 & 19--1399 & 43 & 351.40 (T2) & 32.76 & 0.62 \\
    2016-10-26 09:25:43 & 19--1184 & 46 & 351.40 (T2) & 32.76 & 0.30 \\
    2016-10-28 09:15:52 & 19--1124 & 39 & 355.00 (T3) & 38.30 & 0.35 \\
    2016-10-29 07:10:42 & 19--1124 & 41 & 347.80 (T1) & 33.77 & 1.27 \\
    2016-10-30 07:36:05 & 19--1124 & 39 & 355.00 (T3) & 38.30 & 0.93 \\
    2016-10-30 08:55:42 & 19--1124 & 40 & 347.80 (T1) & 33.77 & 0.78 \\
    2016-11-02 04:23:49 & 19--1124 & 40 & 358.60 (T4) & 30.23 & 0.64 \\
    2016-11-02 05:31:01 & 19--1124 & 40 & 358.60 (T4) & 30.23 & 0.97 \\
    2016-12-17 05:37:41 & 15--460  & 44 & 347.80 (T1) & 33.77 & 0.90 \\
    2016-12-18 05:21:55 & 15--492  & 47 & 347.80 (T1) & 33.77 & 1.29 \\
    2017-04-28 21:51:39 & 15--460  & 39 & 355.00 (T3) & 38.30 & 0.72 \\
    2017-07-03 12:28:53 & 21--2647 & 40 & 358.60 (T4) & 30.23 & 0.24 \\
    2017-07-04 12:40:06 & 21--2647 & 40 & 358.60 (T4) & 30.23 & 0.41 \\
    \enddata
    \tablenotetext{\sharp}{At integration start.}
    \tablenotetext{\dagger}{The number of antenna elements.}
    \end{deluxetable*}

\section{Target} \label{sec:target}

  Among a hundred of $z \gtrsim 8$ candidates from treasury \HST{} programs (e.g., BoRG, CANDELS, CLASH, HFF, HUDF), we carefully selected a bright ($H_{160} < 26$, AB) galaxy candidate with an accurate photometric redshift ($z_\mathrm{phot}$) which is accessible from ALMA ($\delta_{\rm J2000} < +30^{\circ}$) with good atmospheric transmission.  The criteria finally leave \target{} \citep[$H_{160} = 25.92 \pm 0.02$,][]{Laporte15, Infante15}.
  \target{} lies behind the MACS\,J0416.1$-$2403 cluster, one of the \Hubble{} Frontier Fields \citep[HFF,][]{Lotz17}, while the magnification of the LBG is moderate \citep[e.g., magnification factor, $\mu_\mathrm{g} = 1.43 \pm 0.04$,][]{Kawamata16}.  Thanks to the deepest \HST{} and \textit{Spitzer} photometry, the photo-$z$ is well constrained by six independent studies to be
      $z_{\rm phot} = 8.478^{+0.062}_{-0.056}$ \citep{Infante15},
      $8.57^{+0.3}_{-0.4}$ \citep{Laporte15},
      $8.42$ \citep{Laporte16},
      $8.6^{+0.1}_{-0.1}$ \citep{McLeod15},
      8.66 \citep{Castellano16} and
      $8.4^{+0.9}_{-0.9}$ \citep{Kawamata16}.
  The most-likely redshift interval $8.3 < z < 8.7$ can be covered by four tunings of ALMA Band~7, which offer a wide redshift coverage of $\Delta z \approx 0.72$ for \oiii{} 88~$\micron$.


  \begin{figure*}[th]
      \includegraphics[width=\textwidth]{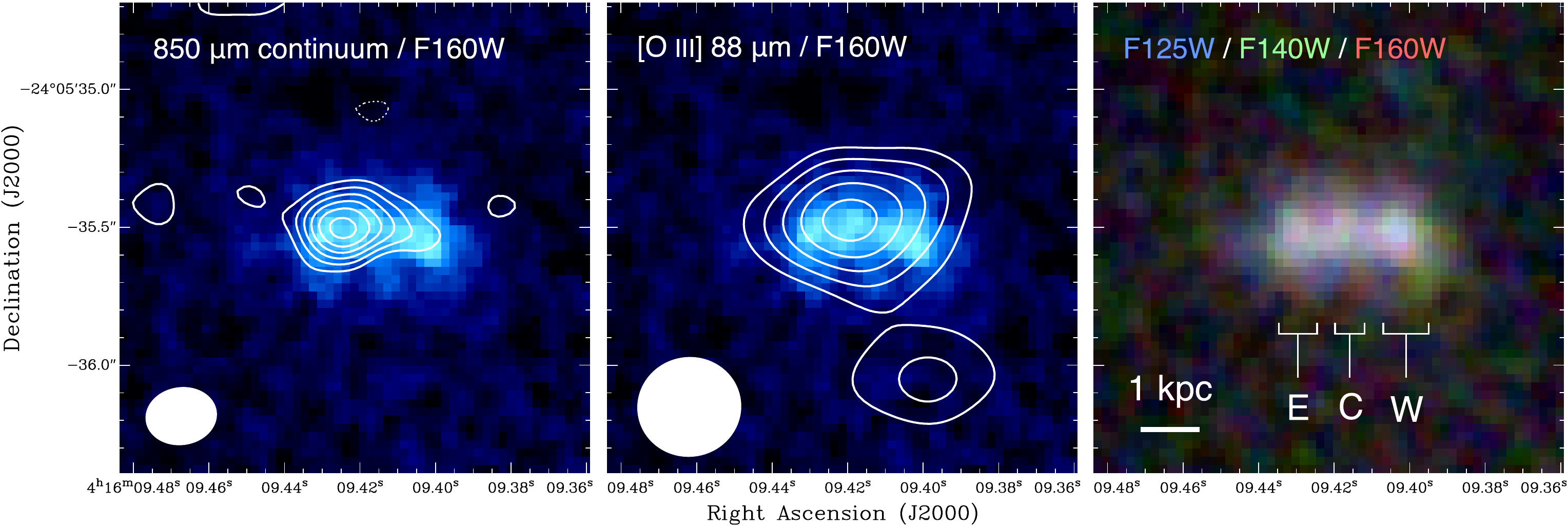}
      \caption{
        (Left) The ALMA 850~$\micron$ continuum image of \target{} (contours) overlaid on the \HST{}/WFC3 near-infrared pseudo-color image in the F160W band. The contours are drawn at $-2\sigma$, $2\sigma$, $3\sigma$, ..., $7\sigma$, where $\sigma = 10.9$~$\mu$Jy~beam$^{-1}$. The negative contour is indicated by the dotted line. The synthesized beam size is indicated at the bottom-left corner.
        (Center) The ALMA \oiii{} 88~$\micron$ integrated intensity image (contours) overlaid on the \HST{}/F160W image. The contours are drawn in the same manner as the Left panel, but $\sigma = 55$~mJy~beam$^{-1}$~km~s$^{-1}$.
        (Right) The false-color \HST{}/WFC3 image taken with F160W (red), F140W (green) and F125W (blue) bands. The letters `E', `C' and `W' denote the positions of the eastern, central, and western clumps seen in the rest-frame UV, respectively. The physical scale of 1~kpc on the image plane is indicated by the bar at the bottom-left corner.
        \label{fig:image}}
  \end{figure*}

\section{Observations} \label{sec:obs}

\subsection{ALMA Observations and Reduction}

    The ALMA observations were carried out from 2016 October to December and 2017 May to July as a Cycle~4 program (program ID: 2016.1.00117.S). The observation log is given in Table~\ref{tab:log}.  Four different tunings were assigned to cover contiguous frequency range between 340.0 and 366.4~GHz. The local oscillators of the Band~7 receivers were tuned at 347.80, 351.40, 355.00, and 358.60 GHz, and we hereafter denote these tunings as T1, T2, T3, and T4, respectively. The correlator was configured with the frequency-division mode, in which the four spectral windows (SPWs) cover 7.5~GHz with respect to the central frequencies with a channel spacing of 7.8125 MHz. The phase tracking center was set to the LBG position, $\rm (\alpha_{J2000}, \delta_{J2000}) = (04^h16^m09\fs4010, -24\arcdeg 05' 35\farcs 470)$, which was determined by the \HST{} observations \citep{Laporte15,Infante15}.  The on-source time was 436~min in total.
    Two quasars, J0348$-$2749 and J0453$-$2807, were used for complex gain calibration.  J0522$-$3627 was used for bandpass calibration.
    Flux was scaled using J0522$-$3627 (for the tunings T2 and T3, $S_\mathrm{850\,\mu m} \simeq 3$--4~Jy) and J0334-4008 (for the tunings T1 and T4, $S_\mathrm{850\,\mu m} \simeq 0.3$~Jy), yielding an absolute accuracy better than 10\%.

    The calibration and flagging were made using a standard pipeline running on \textsc{casa} \citep{McMullin07} version 4.7.2, while manual flagging was needed for some outlier antennas.
    Four tuning data are combined to make the continuum image using the \textsc{casa} task, \texttt{clean}, with the natural weighting.  Note that spectral channels where the \oiii{} line is detected were not used for continuum imaging.  The resulting synthesized beam size in full width at half maximum (FWHM) is $0\farcs 26 \times 0\farcs 21$ (position angle PA = $-82\arcdeg$).  Synthesized beam deconvolution is made down to $2\sigma$.

    Each tuning data set was also imaged to produce a cube with a frequency resolution of 31.25~MHz ($\approx 26$~km~s$^{-1}$) to search for the \oiii{} line. As the data sets were obtained in different array configurations and some SPWs with long baselines may resolve out the emission, we optimally-taper the image with a $0\farcs 35$ Gaussian kernel to maximize the signal-to-noise ratio (SNR) of the emission.
    The resulting beam size and r.m.s.\ noise level measured at 364~GHz are $0\farcs 38 \times 0\farcs 36$ (PA = $-79\arcdeg$) and $\sigma = 0.5$~mJy~beam$^{-1}$, respectively (Figures~\ref{fig:image} and \ref{fig:spectrum2}).

\subsection{ALMA and Hubble Astrometry} \label{sec:astrometry}

    The position of the LBG was originally determined by the HFF \HST{} images which are aligned to the existing CLASH catalogs \citep{Postman12}; the CLASH astrometry was based on Subaru's Suprime-Cam catalogs which are registered onto the Two-Micron All Sky Survey (2MASS) frame.  We find, however, that the optical-to-NIR astrometry does not fully match the International Celestial Reference System (ICRS), on which ALMA relies.  In order to correct the astrometry of the \HST{} images, we use four positions of objects (3 stars and 1 cluster elliptical) accurately measured in the \Gaia{} first data release (DR1) catalog \citep{Gaia16a, Gaia16b}. We also compare the positions of the three quasars used as ALMA phase calibrators, J0348$-$2749, J0453$-$2807 and J0522$-$3627, which are determined by the \Gaia{} DR1 catalog and by phase solutions from our ALMA calibration.
    We find that the relative offsets between the ALMA and \Gaia{} coordinates are typically $< 10$ mas. All of the \HST{} images are corrected for astrometry on the basis of the \Gaia{} coordinates using the \textsc{iraf} \citep{Tody93} task, \texttt{ccsetwcs}, confirming those two frames coincide with each other down to the accuracy of $\lesssim 30$ mas. The resulting centroid of the LBG in the ICRS coordinate is at $\rm (\alpha_{ICRS}, \delta_{ICRS}) = (04^h 16^m 09\fs 415, -24\arcdeg 05' 35\farcs54)$. We hereafter use this coordinate as the formal position of \target{}.

\subsection{X-shooter Observations and Reduction}

  To verify the redshift of \target{} we aimed to detect the redshifted UV emission lines from either Ly$\alpha$, C{\sc\,iv}~$\lambda\lambda$1548,\,1550~\AA{}, C{\sc\,iii}]~$\lambda\lambda$1907,1909~\AA{} or
  O{\sc\,iii}]~$\lambda\lambda$1661,\,1666~\AA{}.

  We observed \target{} with VLT/X-shooter \citep{Vernet11} for a total of 10 hours on target.  The observations were carried out at 10 different nights between 2017 December 8 and 2018 January 21 (Program ID: 0100.A-0529(A), PI: Zackrisson). A blind offset from a neighboring star was used to place the slit on the galaxy using the coordinates from \emph{HST} images. A $1\farcs 2$ wide slit was chosen at optical and near-IR wavelengths in order to capture the rest-frame UV emission from the galaxy, and the slit was aligned along the parallactic angles (between 95$\arcdeg$--100$\arcdeg$ East of North). The observations were carried out in an ABBA nodding mode pattern with $4 \times 900$~s integrations. During the observations, the sky transmission was clear or photometric and the seeing varied between $0\farcs 4$ and $0\farcs 8$, yielding the best possible conditions for detecting faint emission lines.

  The data were reduced with {\sc esorex} scripts \citep{Modigliani10} using adjacent positions in the nodding sequence for sky background subtraction. Corrections for telluric absorption lines were applied from models with {\sc Molecfit} \citep{Kausch15} applied to observations of hot stars following the science integrations. Errors were propagated throughout the data processing stpdf. Since the seeing FWHM was smaller than the slit widths, we measured the spectral resolutions from telluric absorption lines, yielding effectively $R=5600$ in the near-IR spectra. Observations of spectrophotometric standard stars on each of the 10 nights were used to flux calibrate each spectrum before they were co-added.


  \begin{figure*}[ht!]
      \includegraphics[angle=-90,width=1.0\textwidth]{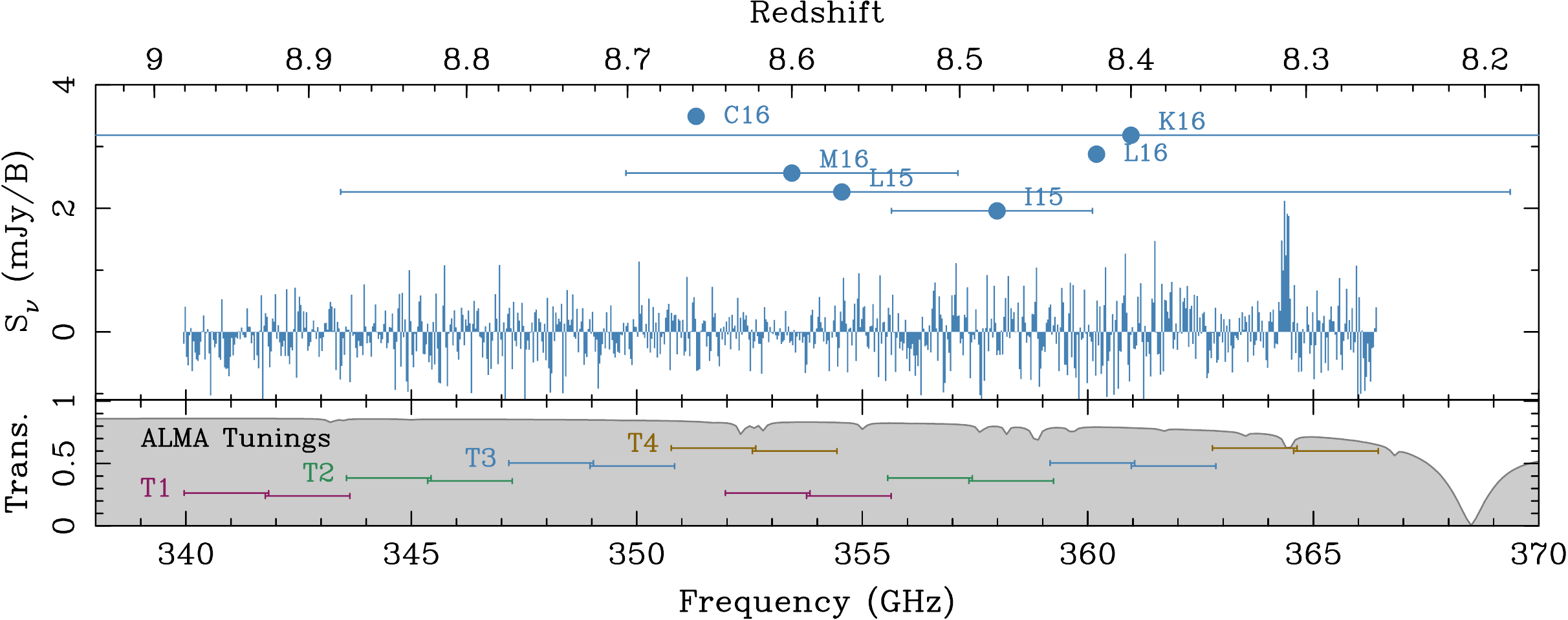}
      \caption{
        The full ALMA spectrum produced by combining four independent tunings
        T1--T4.  The spectrum is continuum-subtracted.  The blue dots with
        error bars represent the photometric redshifts with a 68\% confidence
        interval measured by six independent studies;
        \citet[][denoted as I15]{Infante15},
        \citet[][L15]{Laporte15},
        \citet[][M15]{McLeod15},
        \citet[][L16]{Laporte16},
        \citet[][K16]{Kawamata16} and
        \citet[][C16]{Castellano16}.
        The lower panel shows the atmospheric transmission under a precipitable
        water vapor (PWV) of 0.9~mm, a moderate condition at the ALMA site. The
        horizontal bars show the coverage of the 4 tunings, T1--T4, each of
        which has 4 spectral windows.
        \label{fig:spectrum}}
  \end{figure*}

  \begin{figure}[th!]
      \includegraphics[angle=-90,width=0.47\textwidth]{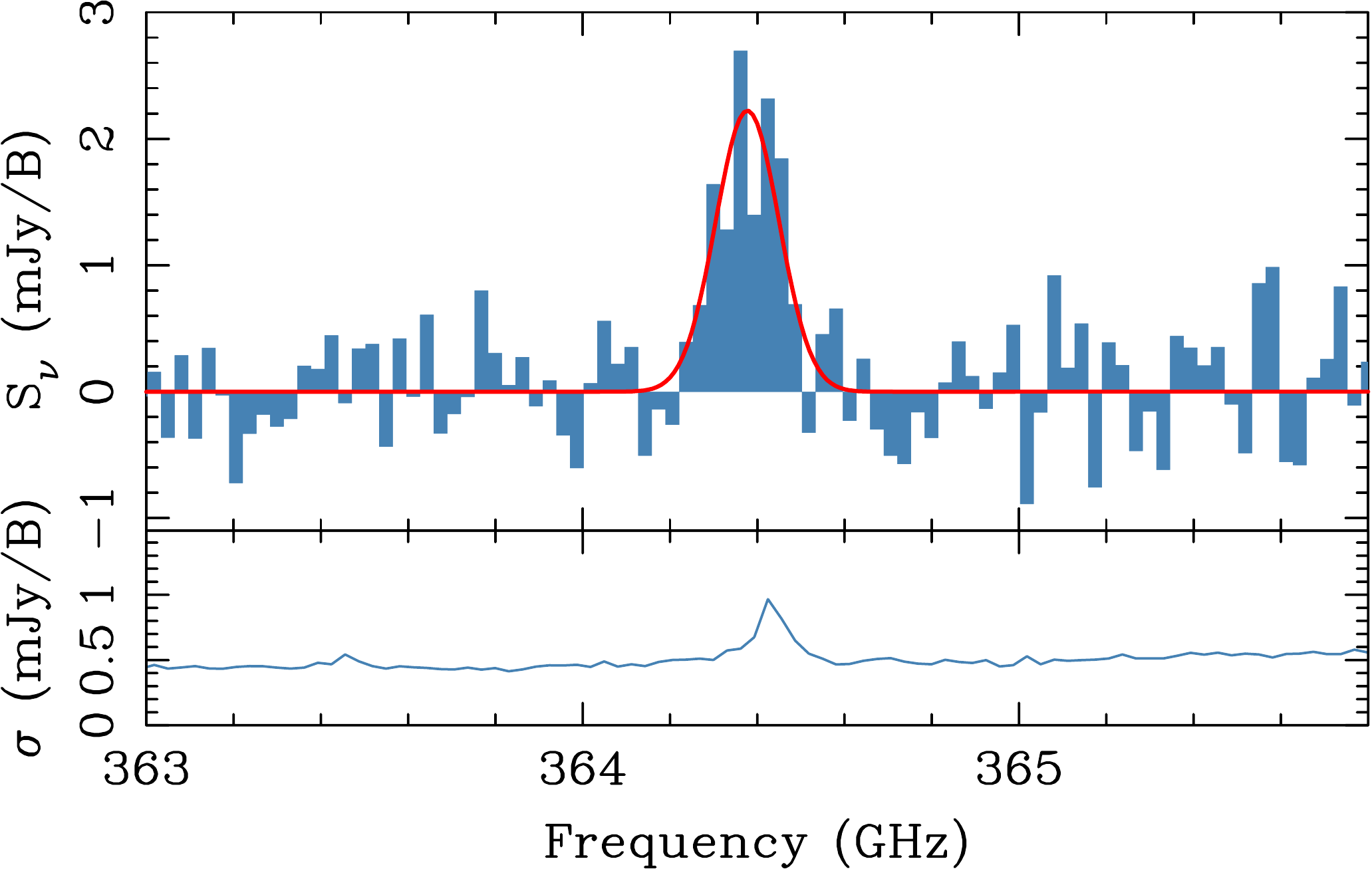}
      \caption{
        The continuum subtracted spectrum showing the \oiii{} line (top) with
        the best-fitting Gaussian function (red curve).  The lower panel shows
        the $1\sigma$ noise level, where an atmospheric absorption line is seen
        close to the \oiii{} line.
        \label{fig:spectrum2}}
  \end{figure}


\section{Results} \label{sec:result}

\subsection{Detection of 850 $\micron$ Dust Continuum}

  We detect 850-$\micron$ (i.e., rest-frame 90~$\micron$) continuum emission at the position of \target{} as shown in Figure~\ref{fig:image} (left). The peak ICRS position is $\rm (\alpha_{ICRS}, \delta_{ICRS})= (04^h16^m09\fs423 \pm 0\fs002,\,-24\arcdeg 05' 35\farcs50 \pm 0\farcs01)$.
  The r.m.s.\ noise level after combining all of the SPWs is $\sigma = 10.9$~$\mu$Jy~beam$^{-1}$, yielding the significance of $7.6\sigma$ on the resulting image.
  The flux density and the image component size deconvolved with the synthesized beam are measured using a \textsc{casa} task \texttt{imfit} with an assumption that the source is 2-dimensional Gaussian and are found to be $S_\mathrm{850\,\mu m} = 137 \pm 26$~$\mu$Jy and $(0\farcs 36 \pm 0\farcs 09) \times (0\farcs 10 \pm 0\farcs 05)$ in FWHM (PA = +84$\arcdeg$), respectively.
  This elongation is not likely due to the cluster magnification but the intrinsic shape of the LBG, because its elongation does not align with the lensing shear direction.  The overall spatial distribution of the 850~$\micron$ continuum is similar to that of the rest-frame UV emission, while the bulk of dust emission is likely to be associated with the eastern `E' knot (or a gap between the `E' and central `C' knots) seen in the \HST{}/WFC3 image (Figure~\ref{fig:image} right).

  We also retrieve a previous 1.14~mm imaging result obtained for the MACS~J0416.1$-$2403 cluster \citep{Gonzalez-Lpez17}. No 1.14 mm emission is found with the $2\sigma$ upper limit of 116~$\mu$Jy. This places an upper limit on the spectral index between 1.14~mm and 850~$\micron$ to be $\alpha > 0.6$ (2$\sigma$), where $\alpha$ is defined such that $S_{\nu} \propto \nu^{\alpha}$.  Despite a relatively weak constraint, this could rule out a low-$z$ interloper with non-thermal emission from an active galactic nucleus, where $\alpha \sim -0.7$ is expected.
  Instead, the spectral index is consistent with dust continuum emission with a temperature of $\gtrsim 30$~K.
  The observed flux density of $S_\mathrm{850\,\mu m} = 137 \pm 26$~$\mu$Jy corresponds to a de-lensed total IR luminosity of
    $\Lir = (1.7 \pm 0.3) \times 10^{11} L_\sun$
          [$(1.1 \pm 0.2) \times 10^{11} L_\sun$]
  and a dust mass of
    $\Mdust = (3.6 \pm 0.7) \times 10^{6} M_\sun$
             [($8.2 \pm 1.6) \times 10^{6} M_\sun$]
  when assuming a dust temperature of $\Tdust = 50$~K (40~K) and the magnification
  factor of $\mu_{\rm g} = 1.43 \pm 0.04$ (see Table~\ref{tab:param} for more details).

      \begin{figure*}[th]
        \begin{center}
          \includegraphics[angle=-90,width=0.65\textwidth]{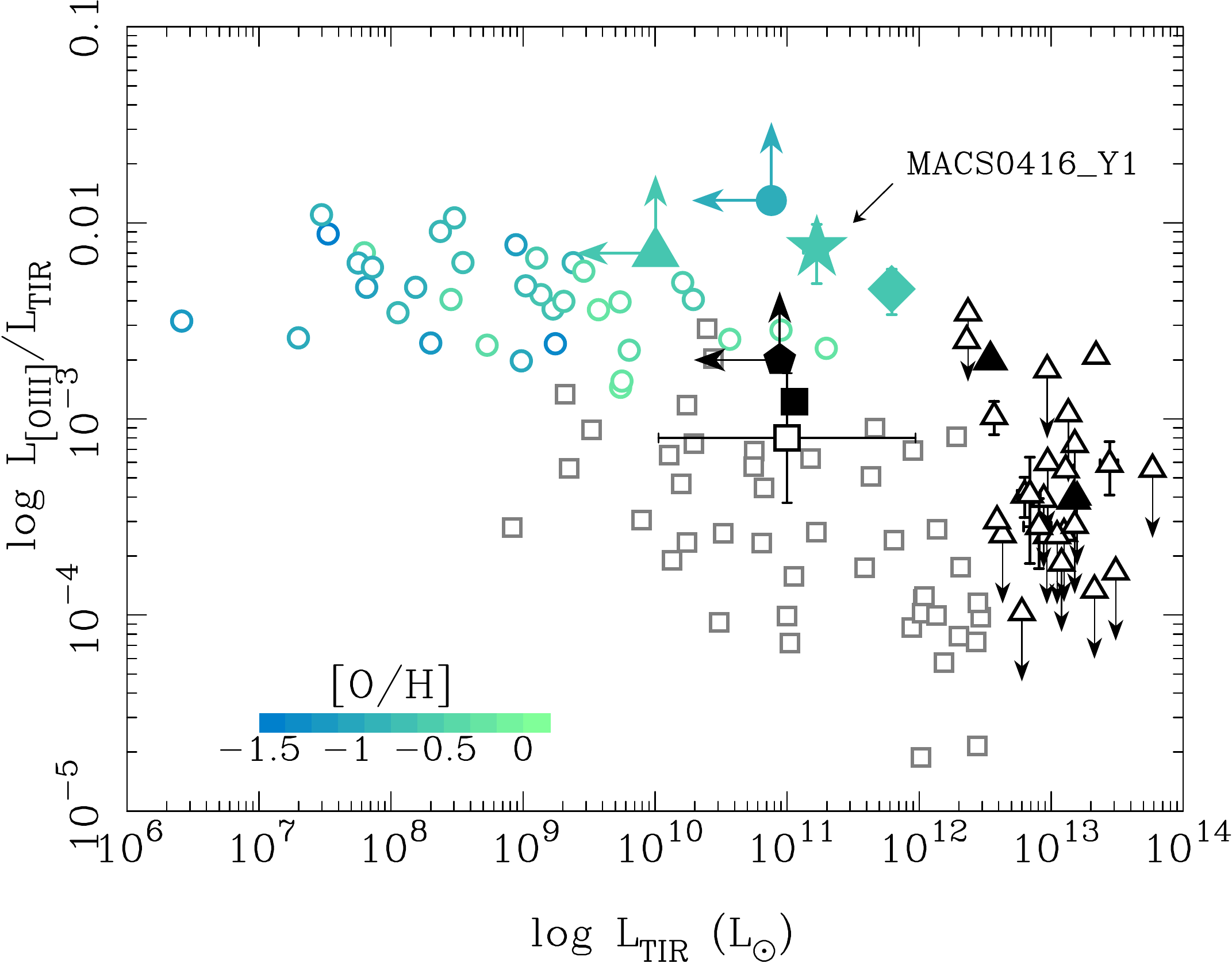}
          \caption{
            The \oiii-to-IR luminosity ratio as a function of IR luminosity. The filled symbols represent the galaxies at $z \gtrsim 7$, i.e.,
              \target{} (five-pointed star),
              SXDF-NB1006-2 \citep[filled circle with two arrows,][]{Inoue16},
              BDF-3299 \citep[filled pentagon,][]{Maiolino15, Carniani17},
              A2744\_YD4 \citep[filled square,][]{Laporte17},
              MACS1149-JD1 \citep[filled triangle with two arrows,][]{Hashimoto18},
              B14-65666 \citep[filled diamond,][]{Hashimoto18b} and
              SPT0311$-$58 E/W \citep[filled triangles,][]{Marrone18}.
              The open symbols are lower-$z$ galaxies;
              local dwarfs \citep[open circles,][]{Madden13, Cormier15},
              the SHINING samples of local star-forming galaxies from normal spirals to ultra-luminous IR galaxies (ULIRGs) \citep[small gray open squares,][]{Herrera-Camus18a, Herrera-Camus18b},
              the median of local spirals \citep[large black open square with the error bar representing 1.5 times the median absolute deviation,][]{Brauher08} and
              $z \sim 2$--4 dusty star-forming galaxies \citep[open triangles,][]{Ferkinhoff10, Ivison10, Valtchanov11, Vishwas18}.
            All of the $z > 7$ galaxies, except for SPT0311$-$58 E/W, are assumed to have the dust temperature of $\Tdust = 50$~K; note that $\Lir$ decreases by a factor of 0.7 if assuming $\Tdust = 40$~K.  The IR luminosities are corrected for magnification, if any.  The blue-to-green color code shown for \target{}, SXDF-NB1006-2, MACS1149-JD1, B14-65666 and local dwarfs indicates the best-fitting oxygen abundances.
            \label{fig:o3deficit}}
        \end{center}
      \end{figure*}

\subsection{Blind Detection of [O\,III] 88 $\micron$}

  At the position of the dust emission, we detect an emission line feature at $364.377 \pm 0.012$~GHz, strongly suggesting the \oiii{} 88~$\micron$ emission line at $z = 8.3118 \pm 0.0003$ (Figure~\ref{fig:spectrum}).  This redshift is slightly lower than, but yet consistent with, the photometric redshifts.  This is a rather common feature seen in LBGs in the reionization era \citep[e.g., MACS1149-JD1][]{Zheng17, Hashimoto18}; the slight offset is likely due to the fact that the largely-neutral interstellar/intergalactic medium attenuates the edge of Lyman break and makes the photo-$z$ estimates higher.
  Figure~\ref{fig:image} (center) shows the integrated intensity image where the \oiii{} line is detected at $6.3\sigma$.  The intensity peak is associated with the `E'--`C' clumps seen in the \HST{}/F160W image (Figure~\ref{fig:image} right).
  The apparent flux is
  $F_\mathrm{[O\,III]} = 0.66 \pm 0.16$~Jy~km~s$^{-1}$, corresponding to the de-lensed luminosity of $L_\mathrm{[O\,III]} = (1.2 \pm 0.3) \times 10^{9} L_{\sun}$.
  The image may barely be resolved and has a beam-deconvolved size of $\sim 0\farcs 5 \times 0\farcs 3$ (PA = $89^{\arcdeg}$) despite a large uncertainty.
  The line width is estimated by a Gaussian fit (Figure~\ref{fig:spectrum2}) and is found to be $\Delta V_\mathrm{[O\,III]} = 141 \pm 21$~km~s$^{-1}$ in FWHM, which is consistent with those predicted for dark halos hosting a bright ($H_{160} \sim 26$) galaxy at $z \sim 8$ in a cosmological hydrodynamic simulation \citep{Shimizu14, Inoue14}.
  The line width is also similar to those found in SXDF-NB1006-2 at $z = 7.212$ \citep[$\Delta V_\mathrm{[O\,III]} \approx 80$~km~s$^{-1}$,][]{Inoue16} and MACS1149-JD1 at $z = 9.110$ \citep[$154 \pm 39$~km~s$^{-1}$,][]{Hashimoto18}, but broader than that of A2744-YD4 \citep[$\Delta V_\mathrm{[O\,III]} \simeq 43$~km~s$^{-1}$,][]{Laporte17}.  The observed quantities are summarized in Table~\ref{tab:param}.

  Figure~\ref{fig:o3deficit} shows the \oiii-to-IR luminosity ratio, $\Loiii/\Lir$, found in local and high-$z$ galaxies as a function of $\Lir$.\footnote{Here we assume $\Tdust = 50$~K for $\Lir$ of $z > 7$ galaxies for which $\Tdust$ is unknown.  The $\Loiii/\Lir$ ratio increases by a factor of 1.5 if assuming $\Tdust = 40$~K.}
  It is known that the $\Loiii/\Lir$ ratios exhibit a possible weak anti-correlation with $\Lir$ as suggested by earlier studies of local galaxies \citep{DeLooze14, Cormier15, Diaz-Santos17}.
  The $\Loiii/\Lir$ of \target{} is estimated to be
  $\Loiii/\Lir \approx 7 \times 10^{-3}$ and is comparable to
  those found in the \Herschel{} Dwarf Galaxy Survey \citep[$\Loiii/\Lir = 5.0^{+16.6}_{-1.5} \times 10^{-3}$,][]{Cormier15}.
  The ratio is as high as those found for $z = 7$--9 UV-selected galaxies,
  such as
  SXDF-NB1006-2 \citep[$\Loiii/\Lir > 1.3 \times 10^{-2}$,][]{Inoue16},
  MACS1149-JD1 \citep[$> 7 \times 10^{-3}$,][]{Hashimoto18},
  B14-65666 \citep[$4.6 \times 10^{-3}$,][]{Hashimoto18b},
  BDF-3299 \citep[$> 2 \times 10^{-3}$,][]{Carniani17}, and
  A2744\_YD4 at $z = 8.38$ \citep[$\sim 1 \times10^{-3}$,][]{Laporte17},
  although SXDF-NB1006-2 and MACS1149-JD1 may have much higher values.
  The ratio of \target{}, however, exhibits a sharp contrast to
  those found in local spirals, ultra-luminous IR galaxies
  \citep[][see the gray open squares in Figure~\ref{fig:o3deficit}]{Herrera-Camus18a, Herrera-Camus18b},
  $z \sim 2$--7 submillimeter galaxies \citep[$\simeq 1 \times 10^{-3}$ or less,][]{Ferkinhoff10, Valtchanov11, Marrone18, Vishwas18},
  and an IR-luminous quasar APM~08279+5255 \citep[$3 \times 10^{-4}$,][]{Ferkinhoff10}.

  \begin{deluxetable*}{ccccccccccc}[ht]
    \tablecaption{The observed quantities of \target{}. \label{tab:param}}
    \tablecolumns{10}
    \tablewidth{0pt}
    \tablehead{
      \colhead{} & \colhead{} &
      \multicolumn{2}{c}{$\Tdust = 40$~K} &
      \colhead{} &
      \multicolumn{2}{c}{$\Tdust = 50$~K} &
      \colhead{} & \colhead{} & \colhead{} & \colhead{} \\
      \cline{3-4}
      \cline{6-7}
      \colhead{$S_\mathrm{850\,\mu m}$} &
      \colhead{FWHM\tablenotemark{$\dagger$}} &
      \colhead{$L_\mathrm{IR}$\tablenotemark{$\sharp$}} &
      \colhead{$M_\mathrm{dust}$\tablenotemark{$\sharp$}} &
      \colhead{} &
      \colhead{$L_\mathrm{IR}$\tablenotemark{$\sharp$}} &
      \colhead{$M_\mathrm{dust}$\tablenotemark{$\sharp$}} &
      \colhead{$F_\mathrm{[O\,III]}$} &
      \colhead{$\Delta V_\mathrm{[O\,III]}$} &
      \colhead{$z$} &
      \colhead{$L_\mathrm{[O\,III]}$\tablenotemark{$\sharp$}} \\
      \colhead{($\mu$Jy)} &
      \colhead{(arcsec)} &
      \colhead{($10^{11} L_\sun$)} &
      \colhead{($10^{6} M_\sun$)} &
      \colhead{} &
      \colhead{($10^{11} L_\sun$)} &
      \colhead{($10^{6} M_\sun$)} &
      \colhead{(Jy km s$^{-1}$)} &
      \colhead{(km s$^{-1}$)} &
      \colhead{} &
      \colhead{($10^{9} L_\sun$)}
    }
    \startdata
      $137 \pm 26$ &
      $0\farcs 36 \times 0\farcs 10$ &
      $1.1 \pm 0.2$ &
      $8.2 \pm 1.6$ &
      &
      $1.7 \pm 0.3$ &
      $3.6 \pm 0.7$ &
      $0.66 \pm 0.16$ & 
      $141 \pm 21$ &
      $8.3118 \pm 0.0003$ &
      $1.2 \pm 0.3$ \\
    \enddata
    \tablecomments{The error represents the 68\% confidence interval.}
    \tablenotetext{\dagger}{
      The beam-deconvolved source size measured for the continuum image.}
    \tablenotetext{\sharp}{
      The value is corrected for lensing magnification of $\mu_\mathrm{g} = 1.43 \pm 0.04$ \citep{Kawamata16}, while the error bar does not include the uncertainty in $\mu_{\rm g}$.  The IR luminosity is derived from a modified blackbody defined in the range of 8--1000~$\micron$.  Note that no Wien correction is applied when the IR luminosity is derived.  The extra heating from the cosmic microwave background is taken into account in deriving the IR luminosity and dust mass according to the formulation by \citet{daCunha13}.  The dust emissivity is assumed such that $\kappa_\mathrm{d}(\nu) = \kappa_\mathrm{d}(850~\micron) (\nu/\nu_0)^{\beta}$,
      where
      $\nu_0 = 353$~GHz,
      $\kappa_\mathrm{d}(850~\micron) = 0.15$~m$^2$~kg$^{-1}$ \citep[e.g.,][]{Weingartner01, Dunne03} and
      $\beta = 1.5$ are the dust absorption coefficient and the emissivity index, respectively.}
  \end{deluxetable*}

\subsection{UV emission line detection limits}

The final 2-dimensional spectrum from X-shooter covering near-IR
wavelengths at 1--2.4~$\micron$ was inspected for emission lines. At
the redshift of \oiii{} 88~$\micron$ from ALMA ($z=8.3118$) no
rest-frame UV emission lines were detected. Furthermore, no emission
lines could be visually identified at other wavelengths. To determine
detection limits, we added artificial emission lines with varying FWHM
at the expected wavelengths and extracted 1-dimensional spectra and
their associated error spectra. To enhance the S/N detection limit, we
binned the data in the spectral dimension by varying factors between 3
and 7 pixels \citep[see][]{Watson15}. To confidently detect an
emission line we require a S/N = 5 detection. Table~\ref{tab:xsh_limits}
summarizes the detection limits for lines with FWHM ranging from 50 to
150~km~s$^{-1}$ for the brighter of the doublet lines.  Since
Ly$\alpha$ is a resonance line, we also compute the detection limit
for a larger width of 250 km~s$^{-1}$.  Typical limits are of the
order of a few times $10^{-18}$~erg~s$^{-1}$~cm$^{-2}$. The reported
limits have not been corrected for lens magnifications.


\begin{deluxetable}{lccc}[hb!]
  \tablecaption{The 5$\sigma$ detection limits of UV emission lines
  in X-shooter data assuming redshift $z=8.3118$.
  \label{tab:xsh_limits}}
  \tablecolumns{4}
  \tablewidth{0pt}
  \tablehead{
    \colhead{} &
    \multicolumn{3}{c}{$5\sigma$ detection limits} \\
    \cline{2-4}
    \colhead{Lines} &
    \colhead{250~km~s$^{-1}$} &
    \colhead{150~km~s$^{-1}$} &
    \colhead{50~km~s$^{-1}$}
  }
  \startdata
    Ly$\alpha$                                     & $< 8.0$  & $< 5.0$ & $< 3.0$ \\
    C{\sc\,iv}   $\lambda\lambda$1548,\,1550 \AA{} & $\cdots$ & $< 4.0$ & $< 1.8$ \\
    O{\sc\,iii}] $\lambda$1666 \AA{}\tablenotemark{$\dag$} & $\cdots$ & $< 5.2$ & $< 3.0$ \\
    C{\sc\,iii}] $\lambda$1907 \AA{}\tablenotemark{$\dag$} & $\cdots$ & $< 6.0$ & $< 2.4$ \\
  \enddata
  \tablecomments{
    The detection limits are measured for line widths (FWHM) of 250, 150 and 50~km~s$^{-1}$. The limits are not corrected for magnification. The flux limits are in units of
    $10^{-18}$~erg~s$^{-1}$~cm$^{-2}$.}
    \tablenotetext{\dagger}{The detection limit for the brighter of the doublet lines.}
\end{deluxetable}

\section{Physical Properties of \target{}} \label{sec:sed}

  The ALMA observations clearly show that \target{} has a substantial amount of dust which exhibits a similar spatial distribution to the rest-frame UV emission on a $\sim$1~kpc scale. This is somewhat surprising because the UV slope is blue ($\beta_\mathrm{UV} \approx -2$) and earlier studies have actually suggested small dust extinction with $A_V \lesssim 0.4$ \citep[e.g.,][]{Laporte15}.
  Furthermore, the \Spitzer{}/IRAC photometry shows a red color in the rest-frame optical ($[3.6]-[4.5] > 0.38$; see Table~\ref{tab:photometry}). The attribution includes (i) the stellar population with the Balmer break at $\lambda_\mathrm{obs} \approx 3.7~\micron$ and (ii) a substantial contribution of the optical \oiii{} $\lambda\lambda4959,\,5007$~\AA{} lines to the 4.5~$\micron$ band \citep[e.g.,][]{Labbe13, Smit15}.  The former case is expected for a relatively-evolved stellar component, while the latter requires a much younger stellar population where OB stars are dominant in luminosity.


  \begin{deluxetable*}{ccccc}[ht!]
    \tablecaption{The photometric data of \target{}.\label{tab:photometry}}
    \tablecolumns{5}
    \tablewidth{0pt}
    \tablehead{
      \colhead{} &
      \colhead{Wavelength} &
      \colhead{AB magnitude\tablenotemark{$\dagger$}} &
      \multicolumn{2}{c}{Flux density\tablenotemark{$\ddagger$}} \\
      \cline{4-5}
      \colhead{Instrument} &
      \colhead{($\micron$)} &
      \colhead{(mag)} &
      \colhead{Value} &
      \colhead{Unit}
    }
    \startdata
      \textit{HST}/F435W & 0.431 & $>30.07$ & $< 3.40$ & nJy  \\
      \textit{HST}/F606W & 0.589 & $>30.40$ & $< 2.51$ & nJy  \\
      \textit{HST}/F814W & 0.811 & $>30.32$ & $< 2.70$ & nJy  \\
      \textit{HST}/F105W & 1.05  & $>29.83$ & $< 4.25$ & nJy  \\
      \textit{HST}/F125W & 1.25  & $26.41\pm0.07$ & $99^{+7}_{-6}$ & nJy  \\
      \textit{HST}/F140W & 1.40  & $26.08\pm0.05$ & $134^{+6}_{-6}$ & nJy  \\
      \textit{HST}/F160W & 1.55  & $26.04\pm0.05$ & $139^{+7}_{-6}$ & nJy  \\
      VLT/HAWK-I ($K_S$) & 2.152  & $26.37\pm0.39$ & $103^{+44}_{-31}$ & nJy  \\
      \textit{Spitzer}/IRAC (ch1) & 3.6 & $>25.32$ & $< 270$ & nJy  \\
      \textit{Spitzer}/IRAC (ch2) & 4.5 & $24.94 \pm 0.29$ & $384^{+117}_{-90}$ & nJy  \\
      ALMA/Band 7 &  850 & ... & $137\pm 26$ & $\mu$Jy  \\
      ALMA/Band 6 & 1140 & ... & $< 116$ & $\mu$Jy  \\
    \enddata
    \tablecomments{All values are not corrected for cluster lensing magnification.}
    \tablenotetext{\dagger}{The error bars represent the 68\% confidence interval.  The lower limit is given at $2\sigma$, where $\sigma$ is obtained by randomly measuring the sky with PSF diameter apertures.  All of the photometry are obtained in this work; see \citet{Laporte15} (\textit{HST} and \textit{Spitzer}) and \citet{Brammer16} (VLT/HAWK-I) for original imaging data.}
    \tablenotetext{\ddagger}{The error bars represent the 68\% confidence interval. The upper limit is given at $2\sigma$.}
  \end{deluxetable*}

\subsection{Spectral Energy Distribution Model}\label{subsec:SED}

  Here we characterize the SED to investigate the physical properties of \target{} by template fits where stellar populations, UV-to-FIR nebular emission, and dust thermal emission are taken into account.
  We use the photometric data of the 850~$\micron$ continuum and \oiii{} line in addition to the rest-frame UV-to-optical bands (Table~\ref{tab:photometry}) to model the SED of \target{}.
  The model is based on the prescription presented by \citet{Mawatari16} and Mawatari et al.\ in preparation,\footnote{Panchromatic Analysis for Nature of HIgh-$z$ galaxies Tool (PANHIT), \url{http://www.icrr.u-tokyo.ac.jp/~mawatari/PANHIT/PANHIT.html}.} where emission components of a stellar continuum \citep{Bruzual03}, rest-frame UV-to-optical nebular lines \citep{Inoue11}, and dust continuum \citep{Rieke09} are accounted for.
  In addition, we take into account the \oiii{} 88~$\micron$ line \citep{Inoue14} and nebular continuum in the UV-to-optical wavelengths \cite{Inoue11}.
  We use the stellar population synthesis model from \citet{Bruzual03} with the Chabrier initial mass function (IMF) defined in the range of 0.1--100~$M_\sun$ \citep{Chabrier03}.  We assume exponentially declining and rising SFRs expressed as
  \begin{eqnarray}\label{eq:sfh}
    \mathrm{SFR}(\tau_\mathrm{age}) =
      \frac{1}{|\tau_\mathrm{SFH}|}
      \exp{\left(-\frac{\tau_\mathrm{age}}{\tau_\mathrm{SFH}}\right)},
  \end{eqnarray}
  where $\tau_\mathrm{age}$ is the age of the galaxy which ranges from 0.1~Myr to the age of the Universe at $z = 8.312$, $\tau_\mathrm{SFH}$ is the $e$-folding time-scale of SFR and is set to $\pm0.01$, $\pm0.1$, $\pm1$, and $\pm10$~Gyr. The positive and negative time-scales represent declining and rising SFRs, respectively.
  For rising SFRs (i.e., $\tau_\mathrm{SFH} < 0$), we just fix
  $\mathrm{SFR}(\tau_\mathrm{age}) = \frac{1}{|\tau_\mathrm{SFH}|} \exp{(10)}$
  at $\tau_\mathrm{age} > 10\tau_\mathrm{SFH}$ to avoid SFRs diverging.
  We also explore five metallicity values of $Z = 0.0001$, 0.0004, 0.004, 0.008, and 0.02 (ranging 1/200--$1\,Z_{\sun}$).
  Three extinction laws from \citet{Calzetti00}, Small Magellanic Cloud \citep[SMC,][]{Prevot84, Bouchet85} and Milky Way \citep[MW, ][who fit data obtained by \citealt{Seaton79}]{Fitzpatrick86} are employed in the range of $A_V = 0$--5 to account for dust extinction.

  The nebular continuum and lines arising from ionized gas are modeled by relating the Lyman continuum (LyC) photon rate, metallicity and nebular emissivity.
  The ionizing photon rate is determined for each grid of stellar age and metallicity in the stellar population synthesis model.
  Following \citet{Inoue11}, we derive the electron temperature, the nebular continuum emissivity for a unit LyC photon rate, and the nebular line emissivities for a unit H$\beta$ flux, as a function of metallicity. H$\beta$ flux is derived by assuming the case B recombination.  The escape fraction of LyC photons, $f_\mathrm{esc}$, is treated as a free parameter, while that of Ly$\alpha$ photons is fixed to 1.  We accordingly attenuate the flux densities at $\lambda_\mathrm{rest} < 912$~\AA{} by a factor of $f_\mathrm{esc}$.  We ignored LyC absorption by dust for simplicity.
  The intergalactic \textsc{H\,i} transmission model of \citet{Inoue14b} is also adopted with an extrapolation from $z < 6$ to $z = 8.3$.  This simply diminishes the flux below the Ly$\alpha$ wavelength including the Ly$\alpha$ emission line.
  For \oiii{} 88~$\micron$, we use the prescription of \citet{Inoue14}, where the \oiii{} emissivity for a unit SFR
  (i.e., $\Loiii{} / \mathrm{SFR}$) is assumed to be modeled as a function of metallicity. This ($\Loiii{} / \mathrm{SFR}$)-to-metallicity relation is calibrated by \textit{ISO}, \Herschel{} and \textit{AKARI} observations of local dwarfs and spirals \citep{Brauher08, Kawada11, Madden13, DeLooze14, Cormier15}.
  We do not use the X-shooter upper limits on the individual UV line fluxes for the model constraint,
  because the UV line intensities are sensitive to physical properties of ionized media, such as electron density and ionization parameter \citep[e.g.,][]{Inoue11}, which could induce large uncertainties in the UV nebular line emissivities.

  We compute the 5-to-1000 $\micron$ luminosity $L_\mathrm{IR}$ for each stellar model grid and extinction by assuming that $L_\mathrm{IR}$ is equivalent to the luminosity of stellar and nebular emission at 0.01--2.2~$\micron$ absorbed by dust. We then assign an IR-to-millimeter SED for each stellar component and extinction on the basis of $L_\mathrm{IR}$. The shapes of the SEDs were determined for local luminous infrared galaxies (LIRGs) and modeled as a function of $L_\mathrm{IR}$ \citep{Rieke09}.


  \begin{figure*}[t]
    \includegraphics[width=0.48 \textwidth]{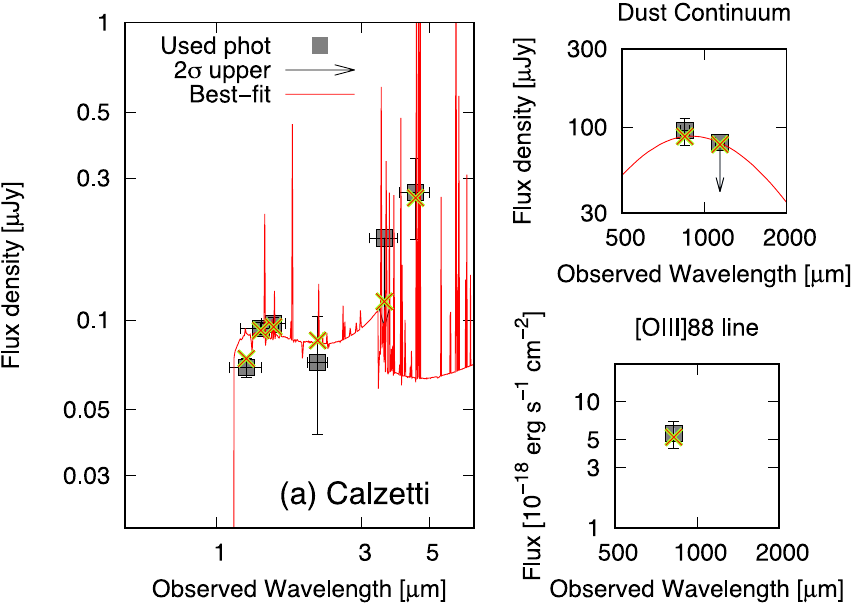}
    \includegraphics[width=0.48 \textwidth]{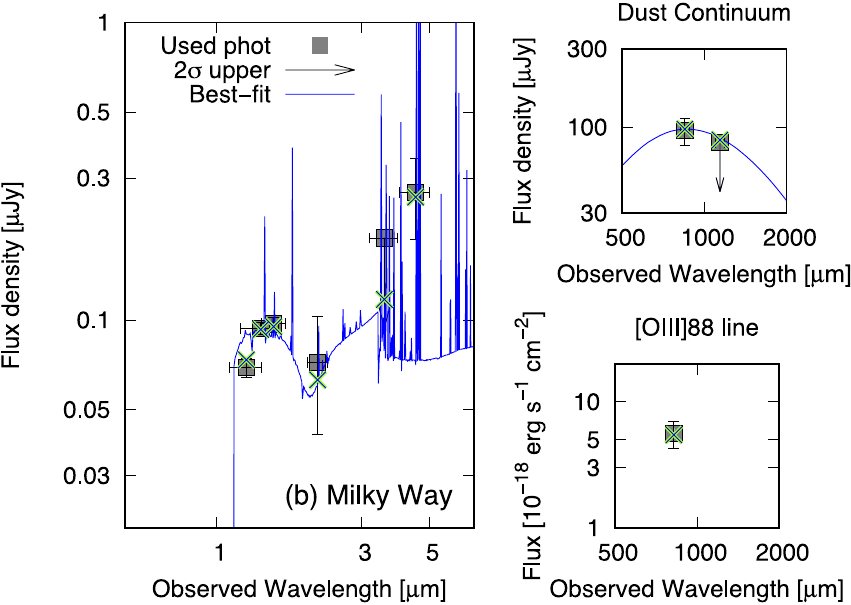}\\
    \includegraphics[width=0.48 \textwidth]{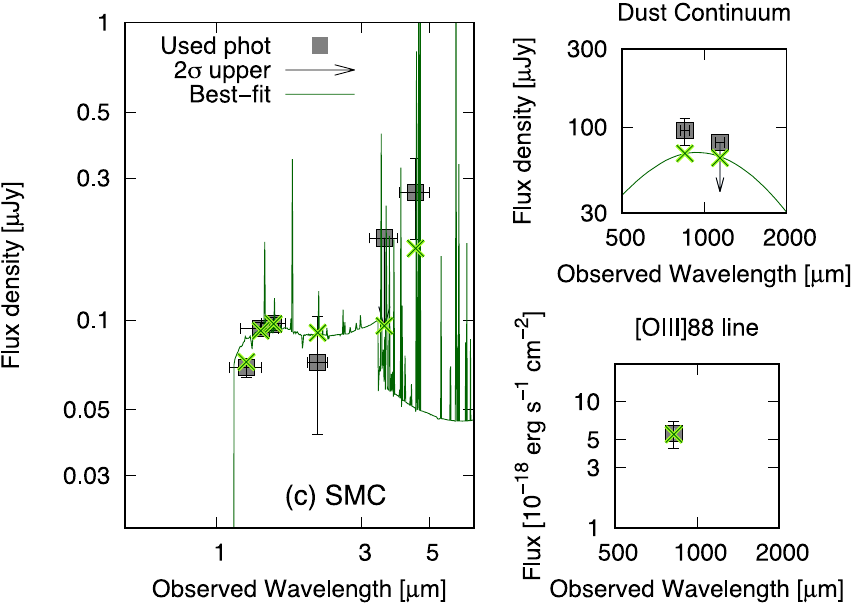}
    \caption{
      The best-fitting spectral energy distributions (SEDs) for three dust extinction laws.
      (a) The SEDs modeled with the Calzetti dust extinction law. The filled and open squares represent the observed photometric data points, while the photometric data at $< 1~\micron$ are not used for SED fits. Those of the rest-frame FIR constraints are shown in the small panels. The solid curve is the best-fitting SEDs. The crosses are flux densities (or flux for the \oiii{} 88~$\micron$ line) predicted from the model.
      (b) The same plot, but the Milky Way dust extinction law is used. The overall trend is the same as the Calzetti case, while the $K_s$ band decrement is well explained by the 2175~\AA{} feature of the extinction law.
      (c) The same plot, but the SMC extinction law is used.
      \label{fig:sed}}
  \end{figure*}

  \begin{deluxetable*}{cccc}[ht!]
      \tablecaption{The best-fitting parameters of the rest-frame ultraviolet to far-infrared spectral energy distribution of \target{}.\label{tab:sedparam}}
  \tablecolumns{2}
  \tablewidth{0pt}
  \tablehead{
      \colhead{} &
      \multicolumn{3}{c}{Extinction law}\\
      \cline{2-4}
      \colhead{Items} &
      \colhead{Calzetti} &
      \colhead{MW} &
      \colhead{SMC}
  }
  \startdata
    $\chi^2$
        & 7.1
        & 6.8
        & 8.4 \\
    Degree of freedom
        & 3
        & 3
        & 3 \\
    Dust attenuation $A_V$ (mag)
        & $0.50 ^{+0.09}_{-0.13}$
        & $0.50 ^{+0.08}_{-0.16}$
        & $0.20 ^{+0.10}_{-0.08}$ \\
    Age $\tau_\mathrm{age}$ (Myr)
        & $3.5  ^{+0.7}_{-2.3}$
        & $4.2  ^{+0.3}_{-3.0}$
        & $2.0  ^{+1.6}_{-0.6}$ \\
    SFH $\tau_\mathrm{SFH}^{-1}$ (Gyr$^{-1}$)\tablenotemark{$^{\sharp}$}
        & $100  ^{+0}_{-200}$
        & $-10  ^{+110}_{-90}$
        & $100  ^{+0}_{-200}$ \\
    Metallicity $Z$ ($Z_{\sun}$)
        & $0.20 ^{+0.16}_{-0.18}$
        & $0.20 ^{+0.17}_{-0.18}$
        & $0.20 ^{+0.54}_{-0.17}$ \\
    LyC escape fraction $f_\mathrm{esc}$
        & $0.00^{+0.19}_{-0.00}$
        & $0.00^{+0.14}_{-0.00}$
        & $0.40^{+0.21}_{-0.40}$ \\
    Stellar mass $M_\mathrm{star}$ ($10^{8} M_\sun$)\tablenotemark{$^{\dagger}$}
        & $2.4 ^{+0.7}_{-0.1}$
        & $2.4 ^{+0.6}_{-0.3}$
        & $2.2 ^{+0.5}_{-0.2}$ \\
    SFR ($M_\sun$ yr$^{-1}$)\tablenotemark{$^{\dagger}$}
        & $57  ^{+175}_{-0.2}$
        & $60  ^{+168}_{-2}$
        & $100 ^{+56}_{-33}$ \\
    $L_\mathrm{IR}$ ($10^{11} L_{\sun}$)\tablenotemark{$^{\dagger}$}
        & $1.5^{+0.2}_{-0.3}$
        & $1.6^{+0.1}_{-0.4}$
        & $1.2^{+0.5}_{-0.2}$ \\
  \enddata
  \tablecomments{The error bars represent the 68$\%$ confidence interval
    estimated from probability distribution functions (PDFs) on the basis of a Monte Carlo technique following the prescription presented by \citet{Hashimoto18}.
    The probability distributions for the fitting parameters are presented in Appendix~\ref{sec:appendix}.
  }
  \tablenotetext{\dagger}{The value is corrected for lensing magnification with $\mu_\mathrm{g} = 1.43 \pm 0.04$ \citep{Kawamata16}, while the error bar does not include the uncertainty in $\mu_\mathrm{g}$.}
  \tablenotetext{\sharp}{$\tau_\mathrm{SFH}^{-1} = 0$, $\tau_\mathrm{SFH}^{-1} > 0$ and $\tau_\mathrm{SFH}^{-1} < 0$ represent constant, exponentially-declining and rising star-formation histories as defined in Equation~\ref{eq:sfh}.}
  \end{deluxetable*}

\subsection{Results}\label{subsec:SEDresults}

  The results are shown in Figure~\ref{fig:sed} and Table~\ref{tab:sedparam}\footnote{See also Appendix~\ref{sec:appendix} for the probability distributions for the SED parameters.}. One of the important outcomes is that there exist solutions which reasonably explain the large amount of dust coexisting with the young stellar components.
  Regardless of the extinction law, the SED fits favor a young, high-SFR solution, where large equivalent widths of the enhanced \oiii{} $\lambda\lambda$4959,\,5007~\AA{} and H$\beta$ lines contribute to the $[3.6]-[4.5]$ color.  The SFR and age are estimated to be $\approx 60\ M_{\sun}$~yr$^{-1}$ and $\approx 4$~Myr, respectively, suggesting that \target{} is at the onset of a starburst phase.
  The carbonaceous absorption feature of the MW extinction law at $\lambda_\mathrm{rest} \approx 2175$~\AA{} can explain the blue $H-K_S$ color, although the predicted flat UV spectrum assuming the Calzetti or SMC law is not ruled out because of a large uncertainty in $K_S$ band photometry.
  The best-fitting metallicity already reaches $Z \approx 0.2 Z_\sun$ at $z = 8.3$ despite a large uncertainty, suggesting rapid enrichment of heavy elements in the middle of the reionization era.

  This does not, however, explain how the galaxy has \emph{obtained} the large amount of dust, even though the SED model explains the energy budget self-consistently if assuming that the dust \emph{preexists}.
  The dust-to-stellar mass ratio inferred from the SED fits is $M_\mathrm{dust} / M_\mathrm{star} \sim 1 \times 10^{-2}$. This is 1--2 orders of magnitude higher compared to the median value obtained for 29 local dwarf galaxies from the \Herschel{} Dwarf Galaxy Survey \citep{Madden13, Remy-Ruyer15}, which is
  $M_\mathrm{dust} / M_\mathrm{star} = 2^{+12}_{-1.8} \times 10^{-4}$,
  where the error bar represents the 90 percentile.
  The ratio would even be an order of magnitude higher than those of dusty star-forming galaxies \citep[e.g.,][]{daCunha10, Smith12, Clark15, DeVis17}; for instance, a median of $M_\mathrm{dust}/M_\mathrm{star} = 4.2 \times 10^{-3}$ is observed for 1402 250-$\micron$ selected normal star-forming galaxies at $z < 0.5$ from the \Herschel{}-ATLAS survey \citep{Smith12}.
  These facts suggest that the stellar mass of \target{} could significantly be underestimated if only ongoing star-formation is taken into account, implying the presence of a more massive, evolved stellar population.  Note that a very high dust temperature (e.g., $\Tdust \sim 100$~K) could only reduce the dust mass by a factor of $\sim 5$ and does not fully explain the high $M_\mathrm{dust} / M_\mathrm{star}$ ratio.  The evolved stellar component as a result of past star-formation activity is also expected as the origin of the dust mass, which will be discussed in the following section \S~\ref{sec:discussion}.



\begin{figure*}
  \includegraphics[angle=-90, width=1.0 \textwidth]{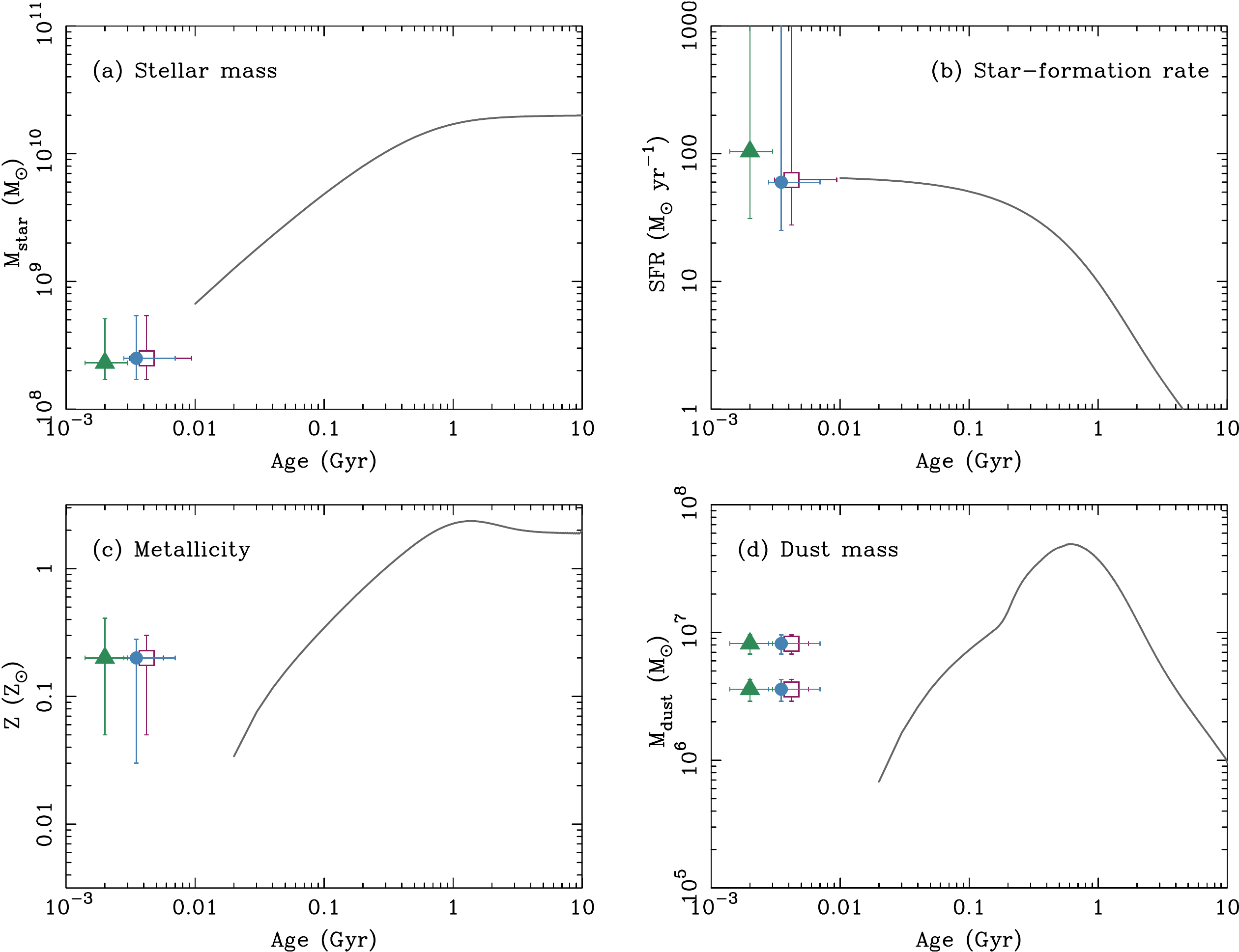}
  \caption{
    The time evolution of (a) stellar mass, (b) SFR, (c) metallicity, and (d) dust mass predicted in a dust formation model \citep{Asano13a, Asano13b, Asano14, Nozawa15} as a function of galaxy age with an initial gas mass of $M_\mathrm{gas} = 2 \times 10^{10} M_{\sun}$ and a star-formation time-scale of $\tau_\mathrm{SFH} = 0.3$~Gyr.
    The open square, filled circle and triangle represent the best-fitting parameters of the physical properties of \target{} estimated by the Calzetti, MW, and SMC extinction laws, respectively (see Table~\ref{tab:sedparam}).
    In panel (d) two de-lensed dust mass estimates for $T_\mathrm{dust} = 50$~K (lower symbols) and 40~K (upper symbols) are plotted.
    \label{fig:grain}}
\end{figure*}

\begin{figure}
  \includegraphics[width=0.47\textwidth]{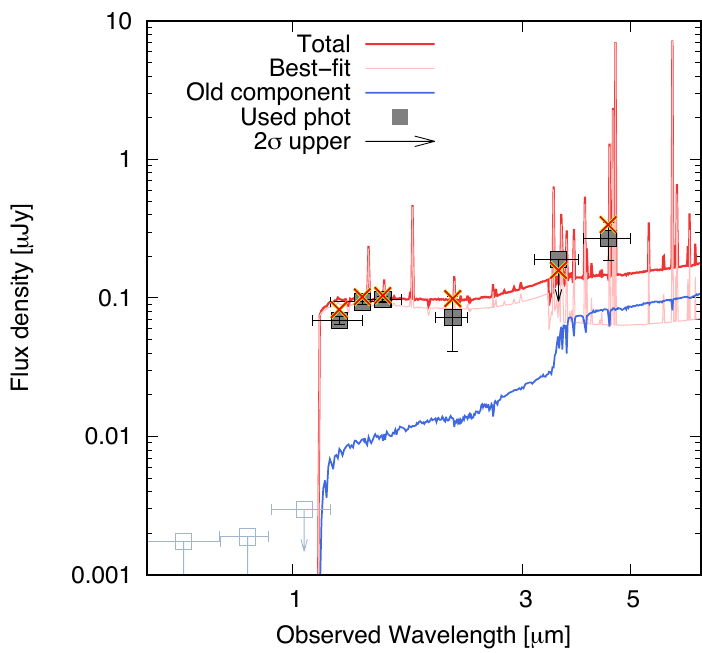}
  \caption{
    The stellar SED model (red curve) in which a mature ($\tau_\mathrm{age} = 300$~Myr) component with $M_\mathrm{star} = 3 \times 10^{9}\,M_{\sun}$ built by constant star-formation lasting for 100~Myr (blue curve) is added to the best-fit SED model (pink curve, \S~\ref{subsec:SEDresults}).  The symbols are the same as those presented in Figure~\ref{fig:sed}, but the crosses are flux densities predicted from the model in which the mature component is added.
    The open squares represent the photometric data at $< 1~\micron$ which are not used for SED fits.
    The extinction law is assumed to be \citet{Calzetti00}.
    \label{fig:youngold}}
\end{figure}


\section{Discussions} \label{sec:discussion}

  The stellar SED analysis presented in \S~\ref{sec:sed} has highlighted
  an extremely-young, star-forming stellar component.
  The solution also explains the IR luminosity if the dust mass preexists, while it needs to be discussed whether the large dust mass can be attained in the very short duration ($\sim 10^{-3}$~Gyr).
  The high $M_\mathrm{dust} / M_\mathrm{star}$ ratio implies the presence of a more massive, evolved stellar component which does not contribute significantly to the rest-frame UV.  However, it is not clear how the evolved component plays a role in dust mass assembly.
  In this section, we discuss how our current understanding of dust formation and evolution reproduces the dust mass observed in \target{}.

\subsection{Dust mass evolution model} \label{sec:dustmodel}


  In the past decade, substantial amounts of dust ranging from $10^6$ to $10^8 M_{\sun}$ have been identified in the $z > 7$ galaxies and quasars \citep[][]{Venemans12, Venemans17, Watson15, Laporte17, Hashimoto18b}, which poses a challenge to the current theory of dust formation \citep{Michalowski15}.
  In the local Universe, the origins of dust grains are the stellar winds of asymptotic giant branch (AGB) stars and ejecta of type II supernovae (SNe~II), although accretion of gas-phase metal onto the grains should play a dominant role.  In fact, grain growth in dense, metal-polluted ISM is considered to be a major contributor to the dust mass in the MW \citep[e.g.,][]{dwek1998, Zhukovska08, draine2009}.
  Furthermore, processes in diffuse gas, such as dust destruction by SN shocks and shattering by grain collisions, control the effective dust yield per SN and the size distribution \citep{Draine79}.
  In the $z > 7$ Universe, SNe~II ejecta with a typical dust yield per SN of $\gtrsim 0.1 M_{\sun}$ are claimed to be more important because at least some of intermediate and low-mass stars did not reach the AGB phase.
  In fact, large dust masses have been reported in some nearby young SN remnants (SNRs), which could explain the high dust yield \citep[e.g.,][]{Dunne03, Dunne09, Barlow10, Matsuura11, Matsuura15, Indebetouw14, Gomez12, Temim13, DeLooze17}, although the other SNRs typically have much smaller dust masses of $10^{-3}$ to $10^{-2} M_{\sun}$ \citep[][and references therein]{Michalowski15}.
  It is likely that the discrepancy is due to the difference in evolutionary phases of SNRs, in which most of dust grains are destroyed by reverse shocks associated with the SN and do not survive in the later SNR phase \citep[][and references therein]{Micelotta18}. If this is the case, it dramatically reduces the yield of dust grains per SN compared to previous beliefs.
  A top-heavy IMF is also claimed to explain a high dust yield per unit SFR \citep{Gall11}, although this could lead to a higher rate of dust destruction by more frequent SN shocks.
  In recent years, a microscopic process of grain growth in dense ISM has been studied to overcome the problem of explaining the high-$z$ dust reservoirs \citep[e.g.][]{liffman1989, dwek1998, draine2009, jones2011}.
  In these models, accretion of gas-phase metal onto dust grains becomes more efficient in dense molecular clouds as the ISM is metal-polluted by stellar ejecta and eventually increases the dust mass for fixed SFR \citep[e.g.][]{Asano13a, zhukovska2014}, allowing one to require neither extreme dust yield per SN nor unusual IMFs.


  Here we employ a dust formation model developed by \citet{Asano13a, Asano13b, Asano14} and \citet{Nozawa15} to assess if the dust mass observed in \target{} is reproduced in the stellar age of $\approx 4$~Myr.
  In this model, time evolution of masses of stars, ISM, metal and dust are solved by 4 independent equations in which SFR, dust injection into (ejection from) stars, dust destruction by SNe, grain growth in dense ISM are considered.  The formation and destruction of dust grains are assumed to be processed in three phases of ISM;
  warm neutral media (WNM, $T_\mathrm{gas} = 6000$~K, $n_\mathrm{H} = 0.3$~cm$^{-3}$),
  cold neutral media (CNM, $T_\mathrm{gas} = 100$~K, $n_\mathrm{H} = 30$~cm$^{-3}$) and
  molecular clouds (MC, $T_\mathrm{gas} = 25$~K, $n_\mathrm{H} = 300$~cm$^{-3}$).
  We assume the mass fractions of WNM, CNM and MC to be $(\eta_\mathrm{WNM},\,\eta_\mathrm{CNM},\,\eta_\mathrm{MC}) = (0.5,\, 0.3,\, 0.2)$ following \citet{Nozawa15}.
  We use the IMF from \citet{Chabrier03}.
  The star formation time-scale is set to $\tau_\mathrm{SFH}$ = 0.3~Gyr, which is well within the 68\% confidence intervals of the best-fitting $\tau_\mathrm{SFH}$ (Table~\ref{tab:sedparam}).
  The total baryon mass is scaled to
  $M_\mathrm{tot} = 2 \times 10^{10}~M_{\sun}$
  such that the model stellar mass and SFR at the galaxy age of
  4~Myr roughly match the best-fitting values
  $M_\mathrm{star} = 3 \times 10^{8}~M_{\sun}$ and $\mathrm{SFR} = 60~M_{\sun}~\mathrm{yr}^{-1}$,
  respectively (Table~\ref{tab:sedparam}).  Note that except this scaling, we made no parameter optimization nor fine-tuning with respect to the best-fitting values obtained by the SED fits.


  The results are shown in Figure~\ref{fig:grain}.
  Although the model does not compute the physical properties at the earliest ages ($< 0.01$~Gyr), the predicted stellar mass and SFR extrapolated from the dust evolution model with a single episode of star-formation (Figure~\ref{fig:grain}a, b) are well aligned with the results from the SED analysis.  The predicted metallicity is, however, much lower (Figure~\ref{fig:grain}d), suggesting that the ISM fed to ongoing star formation is already metal-polluted by past star formation.

  Furthermore, the model fails to reproduce the dust mass (Figure~\ref{fig:grain}d) if assuming that the observed $\Mdust$ would be produced by the ongoing star formation traced by the rest-frame UV continuum and the \oiii{} 88~$\micron$ line.  At the age of $< 0.1$~Gyr, the predicted metallicity and dust mass increase almost linearly with increasing the cumulative number of SNe II.  At $> 0.1$~Gyr, the ISM is sufficiently metal-polluted and triggers a rapid interstellar growth of dust grains, resulting in non-linear evolution of $\Mdust$ at $\tau_\mathrm{age} \sim \tau_\mathrm{SFH}$ = 0.3~Gyr. The dust mass evolution peaks at $\tau_\mathrm{age} \sim 0.6$~Gyr and gives $\Mdust/\Mstar \approx 3 \times 10^{-3}$ (similar to low-$z$ normal star-forming galaxies, see also \S~\ref{subsec:SEDresults}), which is followed by a $\Mdust$ decrement due to grain consumption for star-formation at $\gtrsim 1$~Gyr.
  The galaxy age of $\approx 4$~Myr is too short to reproduce the observed $\Mdust$, and any reasonable modification of the model cannot explain the dust mass.

  \subsection{Potential co-existence of an evolved stellar component} \label{sec:oldcomponent}

  Obviously, the disagreement discussed above should be mitigated if assuming the presence of an underlying `old' stellar component assembled in a past star-formation activity.
  In what follows, we show that there is at least one solution which reasonably explains the physical properties required to reproduce the observed dust mass without any substantial change in the SED shape.

  As for a single episode of star-formation starting with zero metallicity, the time-evolution of metallicity only depends on the elapsed time since the episode started. From Figure~\ref{fig:grain}c, the past star-formation lasting for $\approx 0.1$~Gyr is necessary to reach $Z \approx 0.2\,Z_{\sun}$.
  The dust evolution model also predicts that an initial gas mass of $1\times 10^{10}\,M_{\sun}$ will produce a dust mass of $\Mdust \approx 5 \times 10^{6}\,M_{\sun}$ in 0.1~Gyr.  This stellar population has a virtually-constant SFR of $\approx 30~M_{\sun}$~yr$^{-1}$ for the duration of 0.1 Gyr and attains a stellar mass of $\Mstar = 3 \times 10^9\,M_{\sun}$.

  This massive, old stellar component does not conflict with the best-fitting SED (\S~\ref{subsec:SEDresults}), if the old component has stopped the star-formation activity at a certain point of time in the past and then has been passively evolving for a time duration comparable to the lifetimes of OB stars ($\gtrsim 0.1$~Gyr).
  Figure~\ref{fig:youngold} shows the predicted SED of the old component with $\Mstar = 3 \times 10^9\,M_{\sun}$ built by constant star-formation starting 0.3~Gyr ago and lasting for 0.1~Gyr, which exhibits the Balmer break due to the lack of OB stars.  The UV continuum of the old component is much fainter than that of the best-fit stellar component presented in \S~\ref{subsec:SEDresults}, suggesting that the addition of the old component does not substantially change the stellar SED in the rest-frame UV-to-optical.  A similar star formation history is advocated to account for an excess in the rest-frame optical observed in MACS1149-JD \citep{Hashimoto18} or the presence of dust in B14-65666 \citep{Hashimoto18b}.  Therefore, it is likely that the mature (the age of $\sim 0.3$~Gyr) stellar population with no or little ongoing star-formation may be the origin of the very early enrichment of metal and dust.


\section{Conclusions} \label{sec:conclusion}

  We report the ALMA detections of the \oiii{} 88~$\micron$ line and the 850~$\micron$ dust continuum emission in the \textit{Y}-dropout LBG \target{} located behind the Frontier Field cluster MACS~J0416.1$-$2403.  Four independent tunings of ALMA were assigned to cover the contiguous frequency range between 340.0 and 366.4~GHz (a bandwidth of 26.4~GHz, corresponding to the redshift interval of $\Delta z = 0.72$ around $z \sim 8.5$), which reveals the spectroscopic redshift of $z = 8.3118 \pm 0.0003$.
  The observed 850~$\micron$ flux of $137 \pm 26$~$\mu$Jy corresponds to the intrinsic IR luminosity of $\Lir = 1.7 \times 10^{11} L_{\sun}$ if assuming the dust temperature of $T_\mathrm{dust} = 50$~K and an emissivity index of $\beta = 1.5$, suggesting the fast assembly of a dust mass of $M_\mathrm{dust} = 4 \times 10^{6} M_{\sun}$ when the age of the Universe was 600~Myr.
  The \oiii{} flux and the de-lensed luminosity are $F_\mathrm{[O\,III]} = 0.66 \pm 0.16$~Jy~km~s$^{-1}$ and $\Loiii{} = (1.2 \pm 0.3) \times 10^9 L_{\sun}$, respectively.
  The inferred \oiii{}-to-IR luminosity ratio of $\approx 1 \times 10^{-3}$ is comparable to those found in local dwarf galaxies,
  even if the uncertainty in dust temperature is taken into account.
  The rest-frame UV-to-FIR SED modeling where the \oiii{} emissivity model is incorporated suggests the presence of a young, but moderately metal-polluted stellar component with $M_\mathrm{star} = 3\times 10^8 M_{\sun}$, $Z = 0.2 Z_{\sun}$, $\tau_\mathrm{age} = 4$~Myr.
  The analytic dust mass evolution model with $\tau_\mathrm{SFH} = 0.3$~Gyr, where interstellar grain growth and destruction are fully accounted for, does not reproduce the metallicity and the dust mass in a galaxy age of $\tau_\mathrm{age} = 4$~Myr, suggesting the presence of a past star-formation episode as the origin of dust.  We show that if a stellar population with past star formation triggered 0.3~Gyr ago and lasting for 0.1~Gyr is taken into account, it reproduces the metallicity and the dust mass without any substantial change in the observed stellar SED.

  Obviously, the ionized and neutral ISM in \target{} are yet to be characterized completely.
  The rest-frame FIR and optical fine-structure lines such as \cii{} 158~$\micron$, \textsc{[O\,i]} 146~$\micron$ and \textsc{[N\,ii]} 122/205~$\micron$, \oiii{} 52~$\micron$ and 4959/5007 \AA{}, in addition to multi-wavelength photometry of the continuum emission, will offer a unique opportunity for the better understanding of the fundamental processes of metal/dust enrichment and star-formation activity in \target{}.
  Future ALMA and \textit{JWST} observations of them will allow this to be investigated further.


\acknowledgments
We acknowledges the anonymous referee for detailed comments. We thank K.\ Nakanishi, F.\ Egusa, R.\ Kawamata, Y.\ Harikane, M.\ Ouchi, P.\ Papadopoulos and M.\ Micha{\l}owski for fruitful suggestions.
This work was supported by NAOJ ALMA Scientific Research Grant Numbers 2018-09B, 2016-01A, and JSPS/MEXT KAKENHI (Nos.~17H06130, 17H04831, 17KK0098, 17H01110, 18H04333, 17K14252, and 17H01110).
EZ acknowledges funding from the Swedish National Space Board.
TTT is supported by Sumitomo Foundation Grant for Basic Science Projects (180923).
This paper makes use of the following ALMA data: ADS/JAO.ALMA \#2016.1.00117, ADS/JAO.ALMA \#2013.1.00999.S.  ALMA is a partnership of ESO (representing its member states), NSF (USA) and NINS (Japan), together with NRC (Canada), MOST and ASIAA (Taiwan), and KASI (Republic of Korea), in cooperation with the Republic of Chile.  The Joint ALMA Observatory is operated by ESO, AUI/NRAO and NAOJ.
This work is also based on observations collected at the European Organisation for Astronomical Research in the Southern Hemisphere under ESO programme 0100.A-0529(A).
Some of the data presented in this paper were obtained from the Mikulski Archive for Space Telescopes (MAST). STScI is operated by the Association of Universities for Research in Astronomy, Inc., under NASA contract NAS5-26555.
This work has made use of data from the European Space Agency (ESA) mission {\it Gaia} (\url{https://www.cosmos.esa.int/gaia}), processed by the {\it Gaia} Data Processing and Analysis Consortium (DPAC, \url{https://www.cosmos.esa.int/web/gaia/dpac/consortium}). Funding for the DPAC has been provided by national institutions, in particular the institutions participating in the {\it Gaia} Multilateral Agreement. IRAF is distributed by the National Optical Astronomy Observatory, which is operated by the Association of Universities for Research in Astronomy (AURA) under a cooperative agreement with the National Science Foundation.


\appendix
\section{The probability distribution functions of SED fitting parameters}\label{sec:appendix}

\begin{figure}
  \includegraphics[width=0.80\textwidth]{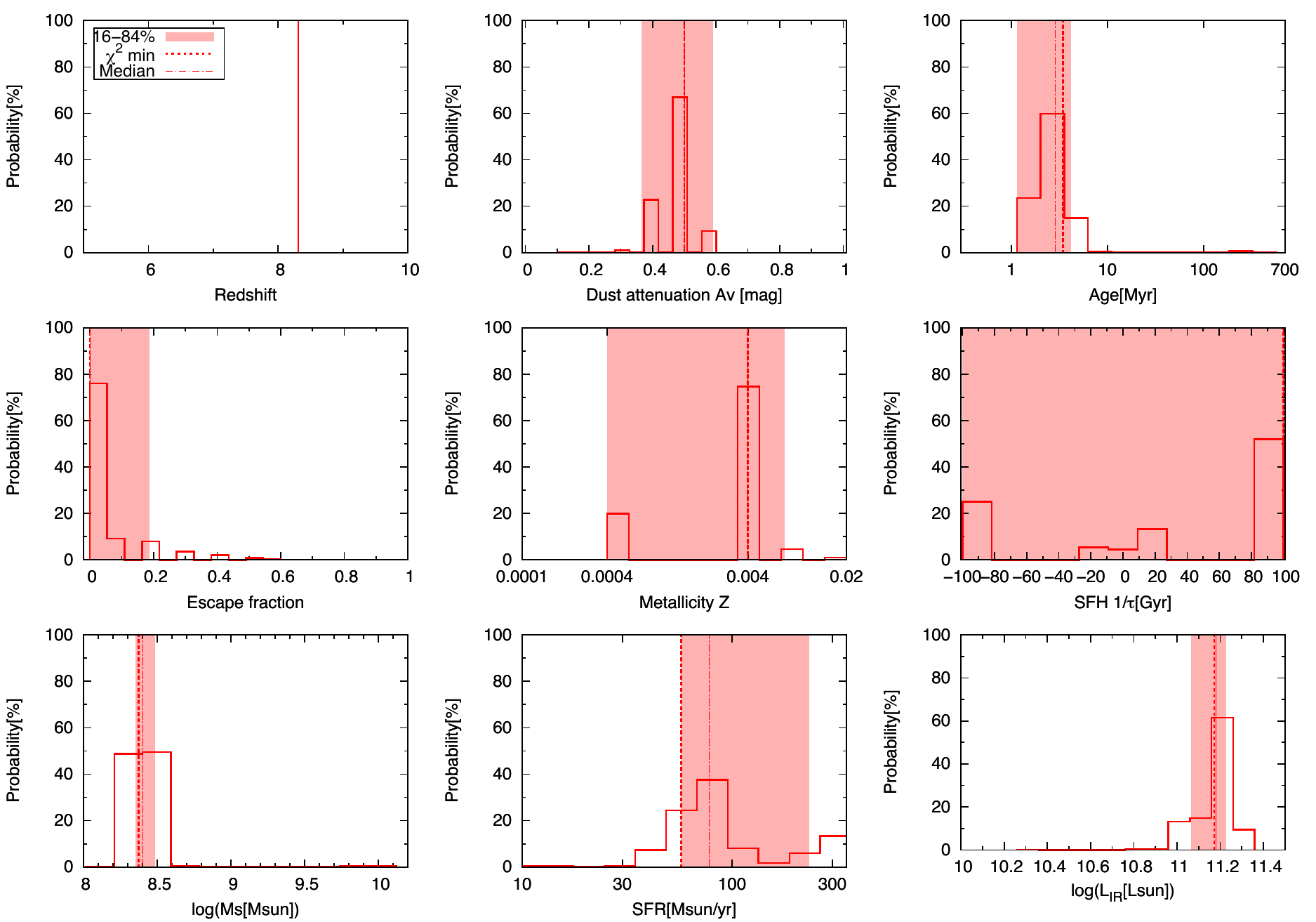}
  \caption{
    The probability distribution functions (PDFs) for the parameters employed in the SED fits presented in \S~\ref{subsec:SEDresults}, such as the dust attenuation $A_V$, the galaxy age $\tau_\mathrm{age}$, the escape fraction of LyC photons $f_\mathrm{esc}$, the metallicity $Z$, the star-formation timescale $\tau_\mathrm{SF}$, the stellar mass $\Mstar$, SFR, and the infrared luminosity $\Lir$. The dotted and dash-dotted lines represent the values which give the minimum-$\chi^2$ and the median of the PDFs, respectively. The shaded areas show the 68\% confidence intervals.
    The PDFs are obtained using the Calzetti extinction law.
    \label{fig:chi2_Calzetti}}
\end{figure}

\begin{figure}
  \includegraphics[width=0.80\textwidth]{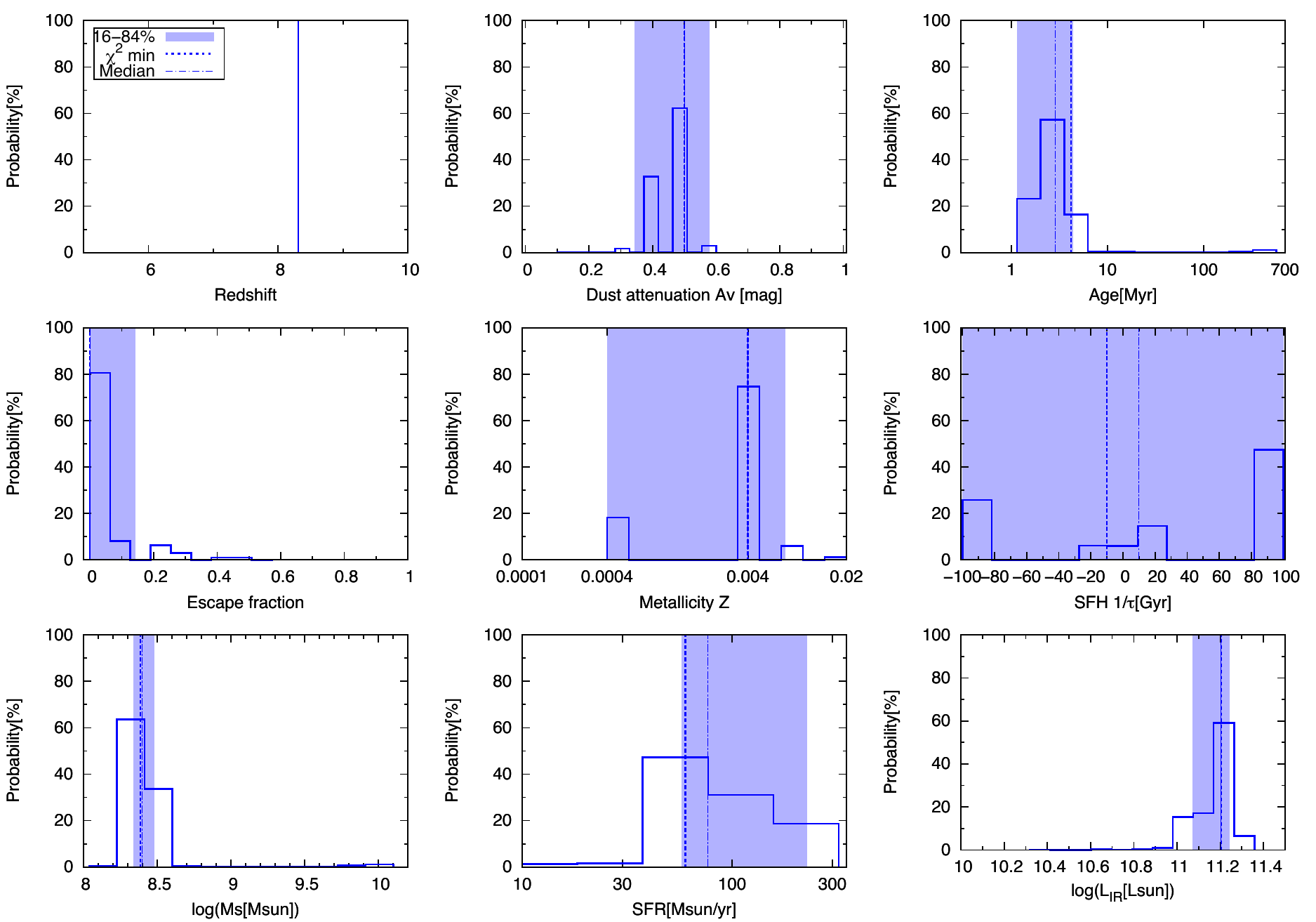}
  \caption{
    The same as Figure~\ref{fig:chi2_Calzetti}, but the probability distributions are obtained using the MW extinction law.
    \label{fig:chi2_MW}}
\end{figure}

\begin{figure}
  \includegraphics[width=0.80\textwidth]{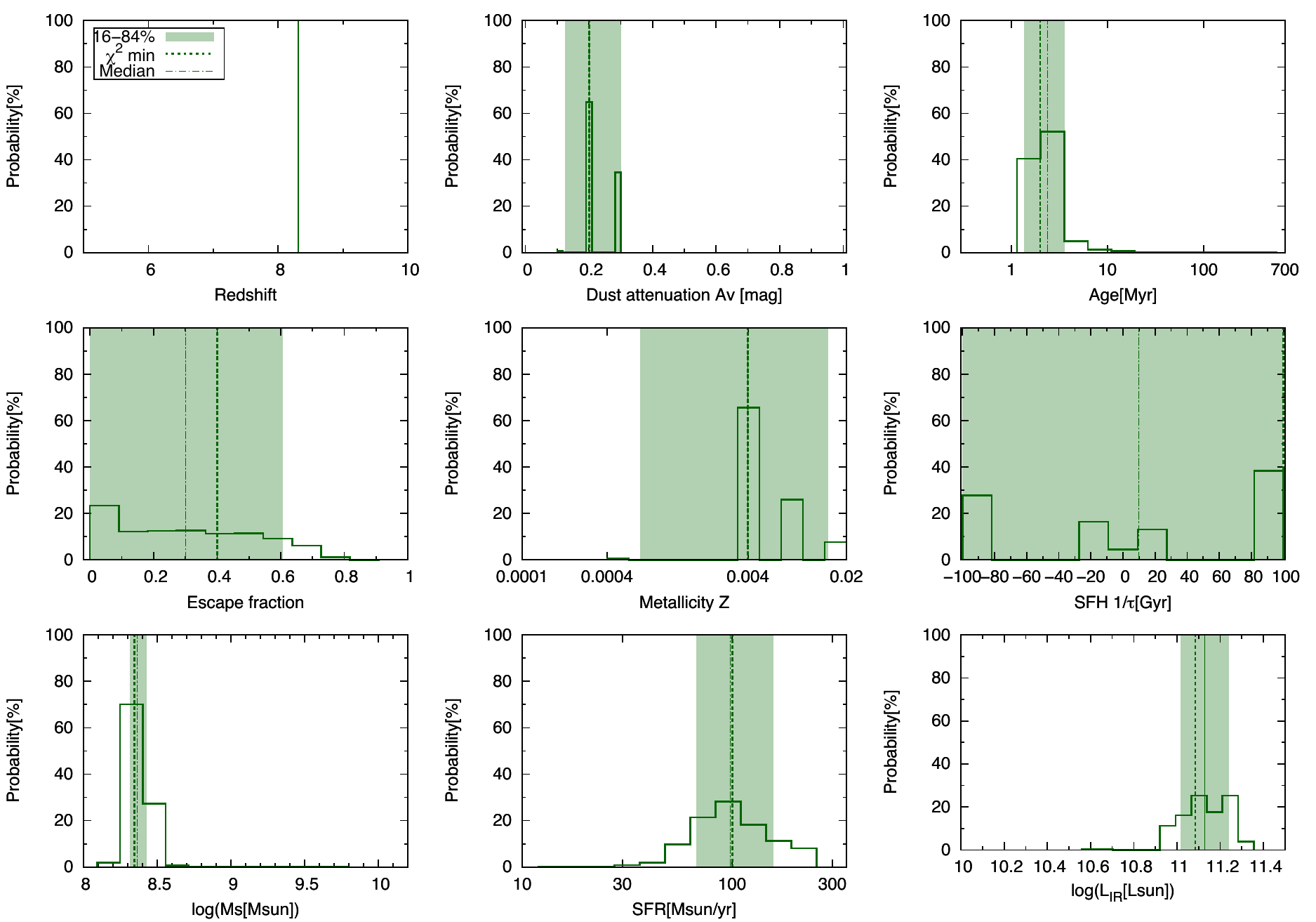}
  \caption{
    The same as Figure~\ref{fig:chi2_Calzetti}, but the probability distributions are obtained using the SMC extinction law.
    \label{fig:chi2_SMC}}
\end{figure}

  Figures~\ref{fig:chi2_Calzetti}, \ref{fig:chi2_MW} and \ref{fig:chi2_SMC} show the probability distribution functions (PDFs) for the SED fitting parameters. See \S~\ref{subsec:SEDresults} for details.

\end{document}